%% file: ms.tex
\documentclass[useAMS,usenatbib]{mn2e_mod}
\usepackage{amsmath,amssymb,graphicx,multirow,xspace,color,hyperref,gensymb}
\usepackage{auto-pst-pdf}

\newcommand{\RUNDMC}{{\tt RUN\,DMC}\xspace}
\newcommand{\MERCURY}{{\tt MERCURY}\xspace}
\newcommand{\SWARMNG}{{\tt Swarm-NG}\xspace}

\newcommand{\B}{\ensuremath{\emph{b}}\xspace}
\newcommand{\C}{\ensuremath{\emph{c}}\xspace}
\newcommand{\D}{\ensuremath{\emph{d}}\xspace}
\newcommand{\E}{\ensuremath{\emph{e}}\xspace}
\newcommand{\F}{\ensuremath{\emph{f}}\xspace}

\newcommand{\Mstar}{\ensuremath{M_{\star}}\xspace} 
\newcommand{\Msol}{\ensuremath{M_{\odot}}\xspace} 
\newcommand{\jitter}{\ensuremath{\sigma_{jit}}\xspace}

\newcommand{\Min}{\ensuremath{\mathcal{M}_\textup{3in}}\xspace}
\newcommand{\Mout}{\ensuremath{\mathcal{M}_\textup{3out}}\xspace} 
\newcommand{\Minstar}{\ensuremath{\mathcal{M}_{\textup{3in}+\sim}}\xspace} 
\newcommand{\Mfour}{\ensuremath{\mathcal{M}_\textup{4}}\xspace} 
\newcommand{\Mfive}{\ensuremath{\mathcal{M}_\textup{5}}\xspace} 

\newcommand{\imutdc}{\ensuremath{\Phi_\textup{dc}}\xspace} 
\newcommand{\imutcb}{\ensuremath{\Phi_\textup{cb}}\xspace} 
\newcommand{\imutbe}{\ensuremath{\Phi_\textup{be}}\xspace} 
\newcommand{\isys}{\ensuremath{i_\textup{sys}}\xspace} 

\newcommand{\setone}{\ensuremath{\textup{Set 1}}\xspace}
\newcommand{\settwo}{\ensuremath{\textup{Set 2}}\xspace}
\newcommand{\setthree}{\ensuremath{\textup{Set 3}}\xspace}

\newcommand{\Laplace}{\ensuremath{\phi_{Laplace}}\xspace}

\newcommand{\chaos}{\ensuremath{\texttt{S}_{Chaos:80}}\xspace}
\newcommand{\synth}{\ensuremath{\texttt{S}_{Reg:20}}\xspace}

\newcommand{\wrel}{\ensuremath{\dot{\omega}_\textup{rel}}\xspace}
\newcommand{\wsec}{\ensuremath{\dot{\omega}_\textup{sec}}\xspace}

\newcommand{\ms}{\ensuremath{\textup{m\,s}^{-1}}\xspace}

\title[3D Resonance in GJ 876]{An Empirically Derived Three-Dimensional Laplace Resonance in the Gliese 876 Planetary System}

\author[B.~Nelson~et~al.]{
Benjamin~E.~Nelson$^{1,2}$,
Paul~M.~Robertson$^{1,2}$,
Matthew~J.~Payne$^{3}$,
Seth~M.~Pritchard$^{4}$,\newauthor
Katherine~M.~Deck$^{5}$, 
Eric~B.~Ford$^{1,2}$,
Jason~T.~Wright$^{1,2}$,
Howard~T.~Isaacson$^{6}$\\
\\
$^{1}$Center for Exoplanets and Habitable Worlds, The Pennsylvania State University, 525 Davey Laboratory, University Park, PA, 16802, USA\\
$^{2}$Department of Astronomy \& Astrophysics, The Pennsylvania State University, 525 Davey Laboratory, University Park, PA 16802, USA\\
$^{3}$Harvard-Smithsonian Center for Astrophysics, 60 Garden Street, Cambridge, MA 02138, USA\\
$^{4}$Department of Physics \& Astronomy, University of Texas San Antonio, UTSA Circle, San Antonio, TX 78249, USA\\
$^{5}$Division of Geological and Planetary Sciences, California Institute of Technology, Pasadena, CA 91101, USA\\
$^{6}$Department of Astronomy, University of California, Berkeley, Berkeley, California 94720, USA\\
}

\begin{document}

\maketitle
\label{firstpage}

\begin{abstract}
We report constraints on the three-dimensional orbital architecture for all four planets known to orbit the nearby M dwarf Gliese 876 based solely on Doppler measurements and demanding long-term orbital stability.
Our dataset incorporates publicly available radial velocities taken with the ELODIE and CORALIE spectrographs, HARPS, and Keck HIRES as well as previously unpublished HIRES velocities.
We first quantitatively assess the validity of the planets thought to orbit GJ 876 by computing the Bayes factors for a variety of different coplanar models using an importance sampling algorithm.
We find that a four-planet model is preferred over a three-planet model.
Next, we apply a Newtonian MCMC algorithm to perform a Bayesian analysis of the planet masses and orbits using an n-body model in three-dimensional space.
Based on the radial velocities alone, we find that a 99\% credible interval provides upper limits on the mutual inclinations for the three resonant planets ($\imutcb<6.20\degree$ for the \C and \B pair and $\imutbe<28.5\degree$ for the \B and \E pair).
Subsequent dynamical integrations of our posterior sample find that the GJ 876 planets must be roughly coplanar ($\imutcb<2.60\degree$ and $\imutbe<7.87\degree$), indicating the amount of planet-planet scattering in the system has been low.
We investigate the distribution of the respective resonant arguments of each planet pair and find that at least one argument for each planet pair and the Laplace argument librate.
The libration amplitudes in our three-dimensional orbital model supports the idea of the outer-three planets having undergone significant past disk migration. 
\end{abstract}


\footnotetext[1]{e-mail: \rm{\url{benelson@psu.edu}.}}

\section{Introduction}

Gliese 876 (=GJ 876) is a 0.37 \Msol M4V star \citep{vonBraun14} hosting four known planets.
This remarkable system represents several milestones: the first detection of a planet around an M-dwarf (GJ 876 b) \citep{Marcy98, Delfosse98}, the first example of multi-planet system orbiting in a mean-motion resonance (MMR) \citep{Marcy01}, the first example of an MMR chain amongst three planets \citep{Rivera10}, and the closest multi-planet exosystem to date (4.689 pc, \citet{vanLeeuwen07}).

The star has a lengthy Doppler (or radial velocity, RV) history spanning two decades and multiple observing sites.
Planet \B was detected contemporaneously by \citet{Marcy98} using the Lick Hamilton Spectrograph and Keck HIRES and \citet{Delfosse98} using the ELODIE and CORALIE spectrographs.
Both estimated a moderately eccentricity for \B ($\sim$0.3) and an orbital period of 61 days for this gas giant from their radial velocity model. 
With more RV observations, \citet{Marcy01} uncovered a second gas giant, \C, orbiting near 30 days.
This planet's RV signature previously masqueraded as a larger eccentricity for planet \B due to the near 2:1 period commensurability of their orbits \citep{Anglada10, Ford11}.
As the Keck RV dataset grew, \citet{Rivera05} revealed a third planet \D orbiting near 1.9 days and was the lowest mass exoplanet around a main-sequence star at the time ($m\,\sin{i}$=5.89$M_\oplus$).
Photometric measurements showed planet \D did not transit \citep{Rivera05,Laughlin05,Shankland06,Kammer14}.
\citet{Correia10} published new HARPS RVs which by themselves could constrain the mutual inclination between planets \C and \B.
Around the same time, \citet{Rivera10} published new Keck RVs which showed an additional RV signal around 124 days, dubbed planet \E.
Numerically integrating their solutions beyond the last observation, the outer three planets (\C, \B, and \E) appear to be in a Laplace resonance, much like the three closest Galilean moons orbiting Jupiter.
Other studies have placed limits on the existence of additional planets and massive companions in the system through observations \citep{Leinert97, Luhman02, Patience02} and considerations of long-term dynamical stability \citep{Jones01, Ji07a, Rivera07, Gerlach12}.

For RV systems, we only observe the component of the planetary induced stellar wobble projected onto our line-of-sight.
Most of the time, there is a degeneracy between the true mass ($m$) and on-sky inclination ($i$), where an edge-on system is $\isys=90\degree$, so we can only place a lower limit on the orbiting companion's mass.
However, if the self-interactions in a multiple planet system are strong, the RV model becomes sensitive to the true masses of the planets, thereby breaking the $m\sin{\isys}$ degeneracy.
There are many RV systems where the true masses can be meaningfully constrained, including HD 200964, 24 Sextantis \citep{Johnson11}, HD 82943 \citep{Tan13}, and other dynamically active systems \citep{Veras10}.

For GJ 876 \C and \B, \citet{Laughlin01b} and \citet{Rivera01} performed self-consistent Newtonian fits and constrain the planetary masses and their on-sky coplanar inclination.
\citet{Rivera05} found an on-sky inclination for the three-planet system ($\isys=50\degree\pm3\degree$), assuming coplanarity.
\citet{Bean09} combined the RVs from \citet{Rivera05} and Hubble Space Telescope astrometry from \citet{Benedict02} and measured the mutual inclination $\Phi$ between \C and \B, where
\begin{equation}
\cos{\imutcb} = \cos{i_c}\cos{i_b}+\sin{i_c}\sin{i_b}\cos\left({\Omega_c-\Omega_b}\right),
\label{eq:mutinc}
\end{equation}
$\Omega$ is the longitude of ascending node, and \C and \B denote the two planets considered.
They find $\imutcb=5.0\pm^{3.9\degree}_{2.3\degree}$.
Based on the HARPS RVs alone, \citet{Correia10} find a lower value ($\imutcb=1.00\degree$).
With their four-planet model, \citet{Rivera10} finds a best fit value $\imutcb=3.7\degree$.
All of the above values are generally consistent with a nearly coplanar system.
When incorporating the effects of correlated noise, \citet{Baluev11} places an upper limit of $\imutcb=5-15\degree$.

Arguably the most studied exoplanet system displaying a MMR, the GJ 876 system has been a prime testbed for dynamical and planet formation theory for the past decade.
The 2:1 MMR of the \C and \B pair is of particular interest, since the libration amplitude of the resonant arguments are a valuable indicator of the system's long-term dynamical stability \citep{Kinoshita01,Ji02,Gozdziewski02,Beauge03,Haghighipour03,Zhou03}.
The most likely mechanism for the planets' current orbital periods and eccentricities is through a combination of planet-planet interactions and disk migration, while most studies have focused on migration through a gas disk \citep{Snellgrove01,Murray02,Lee02,Kley04,Lee04,Kley05,Thommes05,Thommes08b,Lee09,PodlewskaGaca12,Batygin13,Lega13}.
(Semi)-analytic models describing the secular evolution of the system have also been developed \citep{Beauge03,Veras07}.

Other studies of the GJ 876 system include the search for debris disks, thermal properties of \D, and interior structure models of \D \citep[][and references therein]{Rivera10}.
More recent analyses include how the star's UV radiation field affects habitability \citep{France12,France13} and the detectability of additional hypothesized planets in the system \citep{Gerlach12}.
Although the system's orbital architecture appears atypical, GJ 876 does fit into a larger portrait of exoplanet systems around M dwarfs, including population statistics of giant planets \citep{Montet14} and the coplanarity of multi-planet systems \citep{BallardJohnson14}.

RV surveys have discovered a couple dozen strongly-interacting multi-planet systems, motivating the need for analysis procedures to incorporate a self-consistent Newtonian model.
\citet{Benelson14a} developed a Newtonian MCMC algorithm which has been successful in analyzing systems that require a large number of model parameters \citep[e.g. 55 Cancri;][]{Benelson14b}.
The GJ 876 observations have a similar observing baseline and require almost as many model parameters, making it compatible with such an algorithm.

In this work, we present a detailed characterization of the orbits and masses of the GJ 876 planets employing a full three-dimensional orbital model used to fit the Doppler observations.
In \S\ref{sec:obs}, we describe the RV observations made with multiple spectrographs (ELODIE, CORALIE, HARPS, and HIRES), including a new set of HIRES measurements.
In \S\ref{sec:methods}, we describe our orbital and observational model and investigate the effects of correlated noise in the observations.
In \S\ref{sec:coplanar}, we report our results for a coplanar orbital model.
Before advancing to a more complex orbital model, we evaluate the evidence for planets \D and \E by computing Bayes factors as described in \S\ref{sec:FML}.
In \S\ref{sec:3d}, we report the results of the fitting and n-body simulations for a three-dimensional orbital model.
In \S\ref{sec:angles}, we investigate the evolution of the resonant angles and statistics regarding their libration amplitude.
In Appendix \ref{appendix}, we test for possible observational biases in these results.
We conclude with a discussion of the key results and the applications of our posterior samples in \S\ref{sec:discussion}.

\section{Observations}
\label{sec:obs}

\begin{table*}
\centering
\caption{New Keck HIRES velocities for GJ 876. }
\label{tbl-rvs}
\begin{tabular}{cccc}
\hline
\hline
BJD-2450000. [days] & Radial Velocity [\ms] & Uncertainty [\ms] \\
\hline
\hline
\input{table_rvs.tex}
\hline
\hline
\end{tabular}

\medskip
Table \ref{tbl-rvs} is presented in its entirety as Supporting Information with the online version of the article. This stub table is shown for guidance regarding its form and content.
\end{table*}

Our dataset includes publicly-available RVs from four different instruments.
We include 46 ELODIE, 40 CORALIE, and 52 HARPS observations \citep{Correia10}.
The inclusion of these data extends our observing baseline to roughly 580 days before the first Keck HIRES observation.
The 162 Keck observations are reduced by the Carnegie Planet Search group \citep{Rivera10}.
These will be referred to as the Carnegie RVs henceforth.

Our analysis also includes 67 additional Doppler measurements from HIRES reduced by the California Planet Search group (Table \ref{tbl-rvs}).
These will be referred to as the California RVs henceforth.
We measured relative RVs of GJ 876 with the HIRES echelle spectrometer \citep{Vogt94} on the 10-m Keck I telescope using standard procedures.
Most observations were made with the B5 decker (3.5 $\times$ 0.86 arcseconds).
Light from the telescope passed through a glass cell of molecular iodine cell heated to 50\degree\,C.
The dense set of molecular absorption lines imprinted on the stellar spectra between 5000--6200 $\textup{\AA}$ provide a robust wavelength scale against which Doppler shifts are measured, as well as strong constraints on the instrumental profile at the time of each observation \citep{Marcy92,Valenti95}.
We also obtained five iodine-free ``template'' spectra using the B1 decker (3.5 $\times$ 0.57 arcseconds).
These spectra were de-convolved using the instrumental profile measured from spectra of rapidly rotating B stars observed immediately before and after through the iodine cell.
We measured high-precision relative RVs using a forward model where the de-convolved stellar spectrum is Doppler shifted, multiplied by the normalized high-resolution iodine transmission spectrum, convolved with an instrumental profile, and matched to the observed spectra using a Levenberg-Marquardt algorithm that minimizes the $\chi^2$ statistic \citep{Butler96}.
In this algorithm, the RV is varied (along with nuisance parameters describing the wavelength scale and instrumental profile) until the $\chi^2$ minimum is reached.
Each RV uncertainty is the weighted error on the mean RV of $\sim$700 spectral chunks that are separately Doppler analyzed.
These uncertainty estimates do not account for potential systematic Doppler shifts from instrumental or stellar effects.


\section{Methods}
\label{sec:methods}

We characterize the masses and orbits of the GJ 876 planets using the GPU version of the Radial velocity Using N-body Differential evolution Markov Chain Monte Carlo code,  \RUNDMC \citep{Benelson14a}, which incorporates the \SWARMNG framework to integrate planetary systems on graphics cards \citep{Dindar13}.
\RUNDMC has analyzed the 55 Cancri planetary system, which required a high-dimensional ($\sim$40) model.
Compared with 55 Cancri, the problem of GJ 876 seems to be similarly challenging but more computationally tractable due to having fewer observations, a shorter observing baseline, a larger integration timestep, and a lower dimensional model (e.g. one less planet to account for).
On the other hand, the presence of extremely strong planet-planet interactions can result in a challenging posterior distribution that could be more difficult to sample from.

Considering the lessons from \citet{Benelson14a} regarding how to explore parameter space efficiently, we set the following algorithmic parameters for \RUNDMC: $n_{chains}=300$, $\sigma_\gamma=0.05$, and MassScaleFactor=1.0.
To accommodate the inner-most planet's 1.9 day orbital period, we set our integration timestep to roughly 17 minutes and use the time-symmetrized Hermite integrator \citep{Kokubo98b}.

We begin by applying \RUNDMC to the Carnegie RVs from \citet{Rivera10} for a four-planet model.
We sampled from the Markov chains after they had burned-in sufficiently.
In \RUNDMC, we specify the orbital parameters at the epoch of the first observation in our dataset.
For the full RV dataset, this happens to be an ELODIE observation.
To generate the initial states of Markov chains for analyzing the full dataset, we start from a posterior sample based on the Carnegie RVs and rewind each planet's argument of periastron ($\omega$) and mean anomaly ($M$) according to the precession rate ($\dot{\varpi}$), orbital periods, and time difference between the first ELODIE and HIRES observations.
\citet{Laughlin05} find the joint line of apses for the \C and \B pair to be precessing at an average rate of $\dot{\varpi}=-41\degree\textup{ yr}^{-1}$.
After burning-in, we randomly sample from our Markov chains to use as our set of initial conditions for long-term stability simulations using the \MERCURY hybrid integrator \citep{Chambers99}.
The solutions vetted for stability are used as initial conditions for more restricted \RUNDMC runs to be explained in \S \ref{sec:3d}.


\subsection{Model Parameters}
\label{sec:modelParameters}
We characterize the system model with a fixed star mass (\Mstar=0.37\Msol; \citet{vonBraun14}), plus each planet's mass ($m$), semi-major axis ($a$), eccentricity ($e$),  inclination ($i$), argument of periastron ($\omega$), longitude of ascending node ($\Omega$), and mean anomaly ($M$) at our chosen epoch (first ELODIE observation) for each planet, plus the RV zero-point offsets ($C$) and jitters (i.e. unmodeled instrumental and astrophysical noise, \jitter) for each observatory.
We report the orbital periods ($P$) based on Kepler's Third Law and each body's $m$ and $a$ based in a Jacobi coordinate system.
Planet masses and semi-major axes can be readily rescaled to account for any updates to the stellar mass.

The GJ 876 planets are well approximated by a coplanar system, i.e. $\Omega=0\degree$ and $i$ is the same for all planets.
Due to the symmetrical nature of a radial velocity system on the sky, a planetary system at $\isys$ is indistinguishable from one at an inclination of $180\degree-\isys$.
So in \S \ref{sec:coplanar}, we restrict the planets to a coplanar system that can take on any value of $0\degree<\isys<90\degree$.
Similarly, the entire system can be rotated about the line-of-sight.
In \S \ref{sec:3d}, the individual $i$ and $\Omega$ for each planet become free parameters in our model that are fit to the observations.
We set $\Omega_d=0\degree$ to ground our coordinate system.


\subsection{Model of Observations}
\label{sec:modelOfObs}
\RUNDMC allows for fitting multiple zero-point offsets and magnitudes of jitter (e.g. combined astrophysical and/or instrumental noise).
HIRES received a CCD upgrade and new Doppler reduction process in August 2004 (JD 2453241.5).
The Carnegie time series was split based on this pre- and post-upgrade era, which was modeled by two zero-point offsets, but the entire dataset is still modeled with one jitter term.
The California RVs also have a separate offset and one jitter term.
The other instruments (ELODIE, CORALIE, HARPS) were each modeled by one RV zero-point offset and jitter term.
In total, we account for six offsets and five jitter parameters.

Our dataset does not include the HST astrometry from \citet{Benedict02} or any informative priors regarding the inclinations of the GJ 876 planets.

\subsection{Magnitude and Timescale for Correlated Noise}
\label{sec:noise}
Our likelihood function described in Equations 4 and 5 of \citet{Benelson14a} assumes that the observational errors are uncorrelated, which may not be a sufficient approximation for high precision RV measurements of stars with significant stellar activity.
\citet{Baluev11} showed that correlated (=red) noise in the RV data could lead to bias or misestimated uncertainty in some of the orbital parameters (e.g., $e_d$). 
Recently, some discoveries of planets orbiting M dwarfs have been shown to be more likely mere artifacts resulting from stellar activity \citep{Robertson14a,Robertson14b}. 
Therefore, we take a closer look at the role of stellar activity in contributing astrophysical noise to RV observations of  GJ 876.
In \S \ref{sec:FML}, we perform a complementary analysis by computing Bayes factors for a finite set of models.

\begin{figure}
\begin{center}
\centerline{\includegraphics[width=0.9\columnwidth]{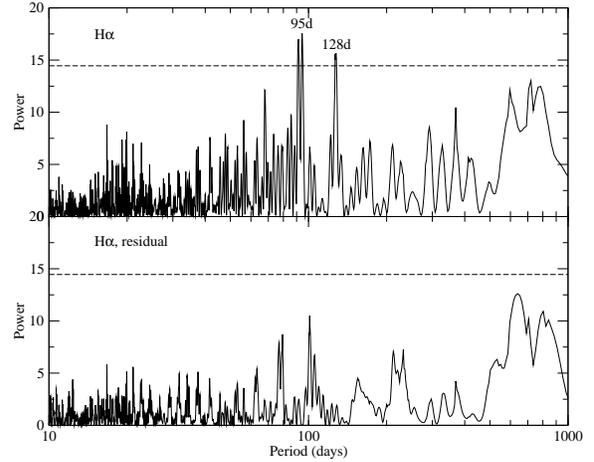}}
\caption{ (Top) Generalized Lomb-Scargle periodogram \citep{Zechmeister09} of the H$\alpha$ stellar activity index for GJ 876, as measured from the publicly available HARPS spectra. We mark the peak at $P = 95$ days--which we adopt as the stellar rotation period--and its 1-year alias at 128 days.
(Bottom) Residual periodogram of H$\alpha$ after modeling and removing a sinusoid at the rotation period.  The dashed lines indicate the power required for a false alarm probability of 1 percent according to Equation 24 of \citet{Zechmeister09}. }

\label{fig:rotation}
\end{center}
\end{figure}

High-precision photometry of GJ 876 \citep{Rivera05} revealed a rotation period of $97 \pm 1$ days.
Examining the variability of the activity-sensitive H$\alpha$ and Na I D absorption lines in the publicly-available HARPS spectra of the star\footnote{Based on data obtained from the ESO Science Archive Facility under request number 129169}, we confirm the rotation period, finding an H$\alpha$ periodicity of $95 \pm 1$ days (Figure \ref{fig:rotation}, top panel).
The appearance of the rotation period in the photometry suggests the presence of starspots, which can affect the measured RVs \citep[e.g.][]{Boisse11, Dumusque14}.
The results from \citet{Baluev11}, \citet{Rivera05}, and our own examination of the spectral activity tracers suggest a complete treatment of activity will yield only marginal improvements in the accuracy of the RV model, and is not necessary to achieve the goals of this study.

\citet{Boisse11} showed that rotating starspots will induce RV periodicities at the stellar rotation period $P_{\textrm{rot}}$ and its integer fractions ($P_{\textrm{rot}}/2$, $P_{\textrm{rot}}/3$, etc.).
Given that none of the planets in the GJ 876 system have periods near the rotation period or its integer fractions, we conclude that rotating starspots have not produced large-amplitude periodic RV signals such as might be misinterpreted as planet candidates.
Although the RV amplitude of the outermost planet \E ($3.5 \ms$) is closest to the amplitudes expected for activity signals produced by a chromospherically quiet star such as GJ 876, we are unaware of any \emph{physical} mechanism that would create an RV signal on this timescale.

Our computed stellar rotation period of $95\pm1$ days beating with one Earth year produces an alias of around 128 days, which is worryingly close to the orbital period of planet \E ($\sim$124 days).
However, subtracting the rotation signal causes the 128-day peak to disappear as well, leading us to suspect that this signal is an alias (Figure \ref{fig:rotation}, bottom panel).
If the 124-day signal is an 1-year alias of the rotation period, then we would also see the 95 day signal in the RV data.
\citet{Rivera10} and \citet{Baluev11} did not find the stellar rotation signal in the RV datasets they analyzed.


\subsection{Computing Bayes Factors for Model Selection}
\label{sec:FML}

When Doppler observations of a system with multiple strongly-interacting planets, the shape of the posterior distribution is often challenging to sample from efficiently.
A general Bayesian approach of performing model comparison is to compute the fully marginalized likelihood, sometimes called the evidence, for each model.
Formally, the evidence is the probability of generating the observed radial velocity dataset $\vec{d}$ assuming some underlying model $\mathcal{M}$ that is parameterized by $\vec{\theta}$,
\begin{equation}
p(\vec{d}|\mathcal{M})=\int p(\vec{d}|\vec{\theta},\mathcal{M})p(\vec{\theta}|\mathcal{M})d\vec{\theta}
\label{eq:FML}
\end{equation}
where $p(\vec{d}|\vec{\theta},\mathcal{M})$ is the likelihood function and $p(\vec{\theta}|\mathcal{M})$  is the prior probability distribution.
While the value of $p(\vec{d}|\mathcal{M})$ is not useful by itself, the ratio of two evidences for two competing models $\mathcal{M}_1$ and $\mathcal{M}_2$, yields the Bayes factor
\begin{equation}
\textup{BF} = \frac{p(\vec{d}|\mathcal{M}_2)}{p(\vec{d}|\mathcal{M}_1)}
\end{equation}
that provides a quantitative measure of which model is preferred and to what degree.

Marginal likelihoods are notoriously difficult to compute.
Integrating Equation \ref{eq:FML} analytically may be possible for some idealized problems with one to a few dimensions, but the model required to describe a 3+ planet system needs roughly 20 or more parameters.
Numerical integration techniques such as Monte Carlo integration become vastly inefficient with increasing dimensionality.

\begin{figure*}
\begin{center}
\centerline{\includegraphics[scale=0.8]{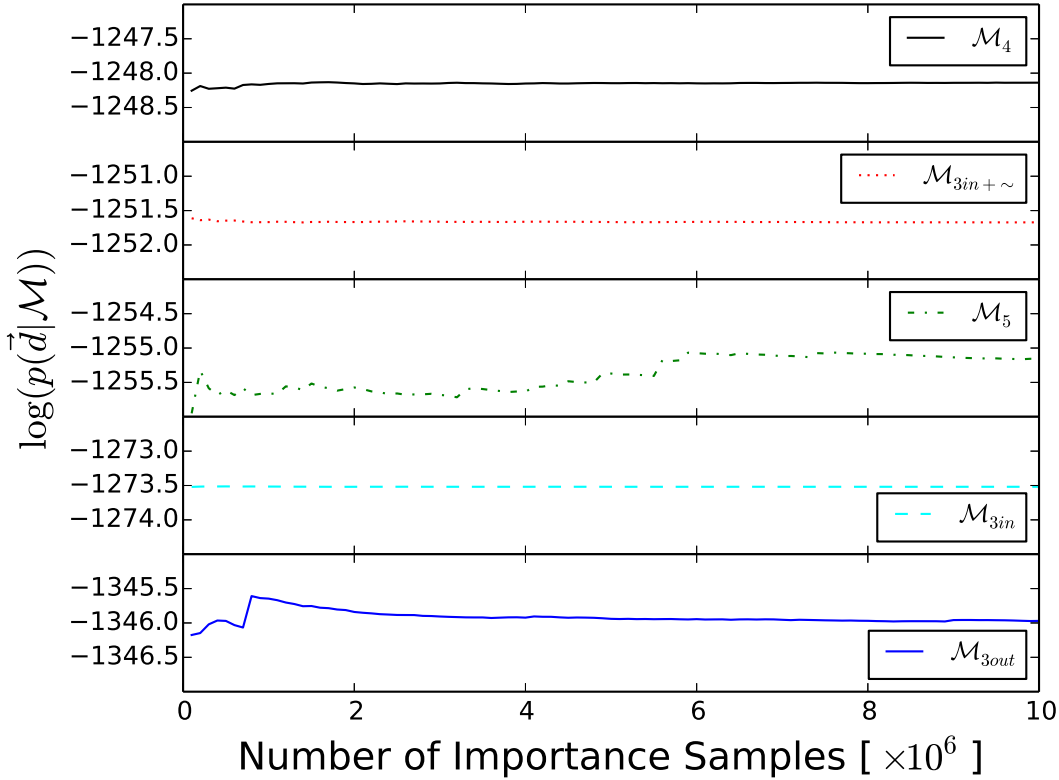}}
\caption{
The estimate of $\log{(p(\vec{d}|\mathcal{M}))}$ (see Equation \ref{eq:FMLfinal}) as a function of number of importance samples for five different models described in \S \ref{sec:coplanar} and Table \ref{tbl-bf}.
Each panel shows the convergence of one model and are ranked from least probable (bottom) to most probable (top): \Mout (solid, blue), \Min (dashed, cyan), \Mfive (dash-dotted, green), \Minstar (dotted, red), and \Mfour (solid, black).
The difference in the minimum and maximum y-values of each panel is kept constant to better compare the variability in $\log(p(\vec{d}|\mathcal{M}))$ for each model we considered.
 }
\label{fig:ISconvergence}
\end{center}
\end{figure*}

\begin{table*}
\centering
\caption{$\chi^2_{eff}$, log evidence [$\log(p(\vec{d}|\mathcal{M})$], and Bayes factors computed for comparison of a finite set of models.}
\begin{tabular}{ccccc}
\hline
\hline
Model & Description & $\chi^2_{eff}$ & $\log(p(\vec{d}|\mathcal{M}))$ & Bayes factor \\
\hline
\hline
\input{table_bf.tex}

\hline
\hline
\end{tabular}
\label{tbl-bf}

\medskip
The Bayes factors in right-most column in Table \ref{tbl-bf} correspond to the ratio of adjacent values of $p(\vec{d}|\mathcal{M})$ in the column immediately to the left.
\end{table*}

Therefore, we use importance sampling, a more general form of Monte Carlo integration, to calculate the integral in Equation \ref{eq:FML} and make this problem computationally tractable.
Following \citet{Ford07}, we sample from a distribution $g(\vec{\theta})$ with a known normalization.
We multiply the numerator and denominator of the integrand in Equation \ref{eq:FML} by $g(\vec{\theta})$,
\begin{equation}
p(\vec{d}|\mathcal{M})=\int \frac{p(\vec{d}|\vec{\theta},\mathcal{M})p(\vec{\theta}|\mathcal{M})}{g(\vec{\theta})} g(\vec{\theta}) d\vec{\theta}.
\label{eq:FMLwithG}
\end{equation}
While the value of $p(\vec{d}|\mathcal{M})$ has not changed, Equation \ref{eq:FMLwithG} can be estimated by drawing $N$ samples from $g(\vec{\theta})$, 
\begin{equation}
\widehat{p(\vec{d}|\mathcal{M})} = \frac{1}{N}\sum\limits_{\vec{\theta}_i \sim g(\vec{\theta})}\frac{p(\vec{d}|\vec{\theta}_i,\mathcal{M})p(\vec{\theta}_i|\mathcal{M})}{g(\vec{\theta}_i)}.
\label{eq:FMLsum}
\end{equation} 

The key aspect of having importance sampling work efficiently is to pick an appropriate $g(\vec{\theta})$.
Assuming our parameter space contains one dominant posterior mode, we choose a multivariate normal with mean vector $\vec{\mu}$ and covariance matrix $\vec{\Sigma}$ for $g(\vec{\theta})$.
For each model considered, we will estimate $\vec{\mu}$ and $\vec{\Sigma}$ from the coplanar MCMC runs described in \S \ref{sec:coplanar}.
Our parameterization for $g(\vec{\theta})$ is $P$, $K$, $e\sin\omega$, $e\cos\omega$, and $\omega+M$ for each planet, the system's orbital inclination $i_{sys}$, and one \jitter for each observatory.
Since we are only interested in computing ratios of $\widehat{p(\vec{d}|\mathcal{M})}$, the priors in zero-point offsets in the calculation will cancel out.

One good strategy with importance sampling is to pick a $g(\vec{\theta})$ that is heavier in the tails than $p(\vec{d}|\vec{\theta},\mathcal{M})p(\vec{\theta}|\mathcal{M})$.
This makes it easier to sample from low probability parts of the posterior distribution and prevents any samples from resulting in extremely large weights.
However, the chance of sampling from the posterior mode will decrease with increasing dimensionality, which could ultimately lead to an estimate of $\widehat{p(\vec{d}|\mathcal{M})}$ that is not efficient.

One way around this is to sample from $g(\vec{\theta})$ within some truncated subspace, $\mathcal{T}$.
This new distribution $g_\mathcal{T}(\vec{\theta})$ is proportional to $g(\vec{\theta})$ inside $\mathcal{T}$ and renormalized to be a proper probability density.
Equation \ref{eq:FMLsum} can be rewritten as
\begin{equation}
f \times \widehat{p(\vec{d}|\mathcal{M})} \approx \frac{1}{N}\sum\limits_{\vec{\theta}_i \sim g_\mathcal{T}(\vec{\theta})}\frac{p(\vec{d}|\vec{\theta}_i,\mathcal{M})p(\vec{\theta}_i|\mathcal{M})}{g_\mathcal{T}(\vec{\theta}_i)}.
\label{eq:FMLtrunc}
\end{equation}
where $f$ is a factor that specifies what fraction of $p(\vec{d}|\vec{\theta},\mathcal{M})p(\vec{\theta}_i|\mathcal{M})$ lies within $\mathcal{T}$.
We can estimate $f$ with an MCMC sample.
By counting what fraction of our posterior samples fell within $\mathcal{T}$, $f_{MCMC}$, we can rearrange Equation \ref{eq:FMLtrunc} to give us $\widehat{p(\vec{d}|\mathcal{M})}$.
\begin{equation}
\widehat{p(\vec{d}|\mathcal{M})} \approx \frac{1}{N \times  f_{MCMC}}\sum\limits_{\vec{\theta}_i \sim g_\mathcal{T}(\vec{\theta})}\frac{p(\vec{d}|\vec{\theta}_i,\mathcal{M})p(\vec{\theta}_i|\mathcal{M})}{g_\mathcal{T}(\vec{\theta}_i)}.
\label{eq:FMLfinal}
\end{equation}
\citet{thesisPC} and \citet{Weinberg13} provide more detailed prescriptions and investigations of this method.

There are two competing effects when choosing the size of our subspace $\mathcal{T}$.
If $\mathcal{T}$ is large (i.e. occupies nearly all of the posterior distribution), then $f_{MCMC}$ approaches 1 and we return to our basic importance sampling algorithm.
If $\mathcal{T}$ occupies a much smaller region, then we are more likely to sample from near the posterior mode, but $f_{MCMC}$ approaches 0, making it difficult to accurately estimate $\widehat{p(\vec{d}|\mathcal{M})}$.
We must carefully choose a $\mathcal{T}$ that will provide a robust estimate of $\widehat{p(\vec{d}|\mathcal{M})}$.
\citet{thesisPC} found that for a three- and four-planet system, truncating the posterior distribution from $-2\sigma$ to $+2\sigma$ was roughly optimal, where $\sigma$ is the standard deviation of each respective model parameter.
We tested several subspace sizes and found $\pm1.5\sigma$ worked well for our problem.

To draw from this distribution, we create a vector $\vec{z}$ whose components are independent draws from a standard normal $\mathcal{N}(0,1)$ truncated at $-1.5\sigma$ and $+1.5\sigma$ for each of our vector components.
Each $\vec{\theta}_i$ is generated as $\vec{\mu} + \vec{A}\vec{z}$, where $\vec{A}$ is the Cholesky decomposition of $\vec{\Sigma}$.
We generate $10^7$ samples from $g(\vec{\theta})$, which provided a robust estimate of $p(\vec{d}|\mathcal{M})$ for each of our five competing models described in \S \ref{sec:coplanar}.
Figure \ref{fig:ISconvergence} shows how well the importance sampling algorithm converged for each model. 


\section{Coplanar Models}
\label{sec:coplanar}

\begin{table*}
\centering
\caption{Estimates of the osculating orbital elements for all the known GJ 876 planets from self-consistent, coplanar dynamical fits. }
\begin{tabular}{ccccc}
\hline
\hline
Parameter & Planet d & Planet c & Planet b & Planet e \\
\hline
\hline
\input{table-cop.tex}
\hline
\hline
\end{tabular}

\medskip
Estimates are computed using 15.9, 50, and 84.1 percentiles. For other details, refer to Section \ref{sec:coplanar}.
\label{tbl-cop}
\end{table*}

\begin{table*}
\centering
\caption{Systematic offset and jitter estimates based on a coplanar orbital model.}
\begin{tabular}{ccc}
\hline
\hline
Dataset & Offset [\ms] & Jitter [\ms] \\
\hline
\hline
\input{table-cop-sys.tex}
\hline
\hline
\end{tabular}

\medskip
Estimates are computed using 15.9, 50, and 84.1 percentiles. For other details, refer to Section \ref{sec:modelOfObs}.
\label{tbl-cop-sys}
\end{table*}

Before attempting to relax the coplanarity constraint, we wish to assess the evidence for all the GJ 876 planets, \E in particular.

We apply \RUNDMC to our cumulative set of 367 RV observations using a coplanar model but allowing for a systematic orbital inclination.
In particular, we consider five different models: one with just the outer-most three planets (\Mout), one with just the inner-most three planets (\Min), one with the inner-most three planets plus a $\sim$124-day sinusoid to mimic either a fourth planet on a circular orbit or a naive model for a stellar activity signal (\Minstar), one with all four planets (\Mfour), and one with a putative fifth planet (\Mfive).

We calculate five fully marginalized likelihoods (four Bayes factors) and summarize the results in Table \ref{tbl-bf}.
Our methodology for computing these probabilities are described in \S \ref{sec:FML}. 

The Bayes factor for \Min/\Mout is $\sim$$10^{31}$, decisively favoring \Min.
This was expected since \Min accounts for the three most significant RV signals in GJ 876.
The Bayes factor for \Minstar/\Min is $\sim$$10^{10}$.
Despite the increased parameterization of \Minstar, this model with four signals is decisively favored over \Min.
The Bayes factor for \Mfour/\Minstar is $\sim$$30$.
The ratio of the evidence for a fully interacting four planet model \Mfour relative to an interacting three planet model with a decoupled $\sim$124-day signal \Minstar is modest.
Regardless, a model with four signals is overwhelmingly preferred over a model with three signals.

Strictly considering the results above, we would not be able to select \Mfour over \Minstar.
Nevertheless, we do not expect this $\sim$124-day candidate to be due to (an alias of) stellar magnetic activity.
\citet{Rivera10} found that after fitting for the inner-three planets to the Carnegie RVs alone, the residual periodogram strongly peaked at the planet \E's period.
The periodogram did not contain a strong signal near the stellar rotation period (see \S \ref{sec:noise}), so we do not suspect the 124-day signal is an alias.

For \Mfive, we initialized \RUNDMC simulations with a hypothetical planet \F with the following properties: $P_f=15.04$ d, $K_f=3 \ms$, $e_f=0.007$, $\omega_f=78.3\degree$, and $M_f=159.8\degree$.
This set of initial conditions was inspired by an additional candidate signal uncovered by \citet{Jenkins14} using a global periodogram method and RVs extracted from HARPS spectra using the HARPS-TERRA pipeline \citep{AngladaButler12}.
One significant difference is that we use an RV signal much closer to the noise level and HARPS RVs based on \citet{Correia10}.
However, this periodicity is suspect since it is close to an integer fraction of the two most prominent signals in the system, \C ($P_c/2$) and \B ($P_b/4$).
This makes our choice of \Mfive a good test for model comparison, since the strength of this signal competes with the extra parameterization needed to model it.
After burning-in, we obtain the following estimates for \Mfive: $P_f=15.07\pm^{0.32}_{0.29}$ d, $K_f=0.42\pm^{0.42}_{0.31} \ms$, and $e_f=0.16\pm^{0.21}_{0.12}$.
Due to the low RV amplitude, $\omega_f$ and $M_f$ took on any value from 0\degree\,to 360\degree\,almost uniformly.
We randomly draw 10,000 posterior samples from our Markov chains for each of these cases and report our values of $\chi^2_{eff}$ in Table \ref{tbl-bf}, along with the results of our Bayesian model comparison tests.
These $\chi^2_{eff}$ values incorporate a penalty term dependent on radial velocity jitter (Equation 5; \citet{Benelson14a}).

We test the coplanar models for short-term orbital stability using the hybrid integrator of \MERCURY \citep{Chambers99}.
A subsample of 1,000 models is integrated for $10^6$ years.
If at any point a planet collides with another body or the semi-major axis of any planet changes by more than 50\% of its original value (i.e. $\left|[a_{final}-a_{initial}]/a_{initial}\right|>0.5$), then the simulation stops and is tagged as being unstable.
This is the default setup for our dynamical integrations unless stated otherwise.

For \Mout, we find 93\% of the posterior samples are stable for the duration of the integration.
The mode of instability for the remaining 7\% was $\left|\triangle a/a\right|$ of planet \E exceeding 0.5.
This typically happened when $e_e>0.14$.
For both \Min and \Minstar, all of the posterior samples are stable for the same integration timespan.
For \Mfour, 99\% of the posterior samples are stable.
For \Mfive, only 18\% of the posterior samples were stable.
77\% of models had $\left|\triangle a/a\right|$ of \E exceeding 0.5, and at least 5\% of models involve a planetary collision.
Some simulations had both instabilities occur.

In \Mfour, both the inner and outer Laplace pair were locked in a 2:1 mean-motion resonance, i.e. at least one resonant argument was librating (for details on resonant arguments, see \S \ref{sec:angles}), for nearly all of our posterior samples.
The Laplace argument was also librating in nearly all of our samples.

\begin{figure}
\centering
\includegraphics[width=0.9\columnwidth]{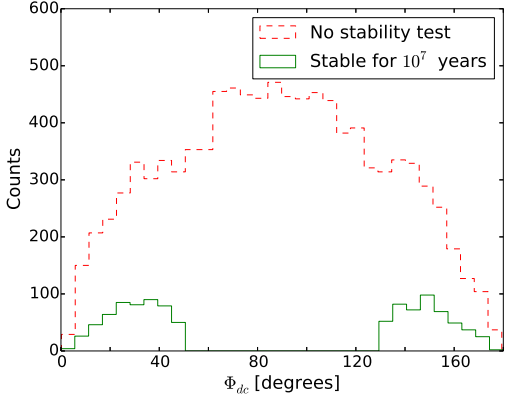}
\caption{ Mutual inclination distribution for planet's \D and \C. The dashed red histogram shows the initial 10,000 posterior samples from \setone before any stability tests are performed. The solid green histogram shows a sample of 1,000 systems from \setthree that were stable for at least $10^7$ years.}
\label{fig:planetDstability}
\end{figure}

\begin{table*}
\centering
\caption{The different orbital parameter constraints applied to \RUNDMC. }
\begin{tabular}{ccccc}
\hline
\hline
Sample set & & Constraints during \RUNDMC & & Then Integrate For... \\
\hline
\hline
\setone & $0\degree\,< \imutdc < 180\degree$ & $0\degree\,< \imutcb < 180\degree$ & $0\degree\,< \imutbe < 180\degree$ & $10^5 \textup{yr}$ \\
\hline
\multirow{2}{*}{\settwo} & $0\degree\,< \imutdc < 60\degree$ & $0\degree\,< \imutcb < 180\degree$ & $0\degree\,< \imutbe < 13\degree$ & $10^6  \textup{yr}$ \\
 & $120\degree\,< \imutdc < 180\degree$ & & & \\
\hline
\multirow{2}{*}{\setthree} & $0\degree\,< \imutdc < 60\degree$ & $0\degree< \imutcb<4\degree$ & $0\degree<\imutbe < 3\sqrt{16-\imutcb^2}\degree$ & $10^7  \textup{yr}$ \\
 & $120\degree\,< \imutdc < 180\degree$ & & & \\
\hline
\hline
\end{tabular}

\medskip
\label{tbl-sets}
Here we show what range of parameter space \RUNDMC is allowed to explore for each Set of simulations. The initial ensemble for a particular Set is generated based on stable regime found from the previous Set.
\end{table*}

\begin{figure*}
\begin{center}
\centerline{\includegraphics[scale=1.]{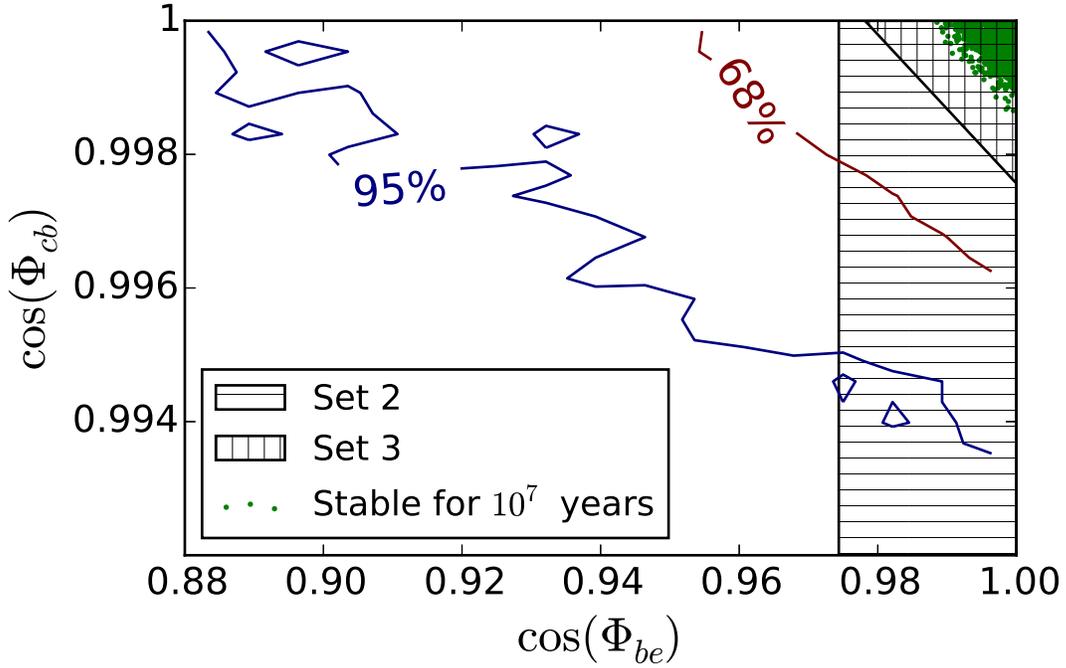}}
\caption{ The joint cosine mutual inclination distribution for planets \C and \B (vertical axis) and \B and \E (horizontal axis).
Contours map the approximate 1-$\sigma$ (68\%) and 2-$\sigma$ (95\%) credible regions for the initial 10,000 posterior samples from \setone before any stability tests are performed.
Horizontal hatch marks indicate the region for obtaining Set 2 and the vertical hatch marks indicate the region for obtaining Set 3, both described in \S \ref{sec:3d} and Table \ref{tbl-sets}.
Green dots show a sample of 1,000 systems from \setthree that were stable for at least $10^7$ years. }
\label{fig:planetEstability}
\end{center}
\end{figure*}

Based on these previous works and the results of \S \ref{sec:noise} and \S \ref{sec:FML}, we conclude that \Mfour is the best model to generate these observations to date.
With a four-planet model decisively chosen, we report estimates of orbital parameters and instrumental properties based on a coplanar model in Tables \ref{tbl-cop} and \ref{tbl-cop-sys} respectively.

Before moving onto our three-dimensional model, we consider the significance of the relativistic precession rate \wrel for planet \D, since it is estimated to have a moderate eccentricity ($e_d\,\sim$0.1) and is close to its host star.
While the rapid resonant interactions amongst the outer-three planets dominate their orbital evolution, the inner-most planet is mostly dynamically decoupled and undergoes secular perturbations.
A value of \wrel comparable to the secular precession rate \wsec tends to aggravate an instability \citep{Ford00a}, while an \wrel that is much more rapid than \wsec could quench secular effects \citep{Adams06b,Adams06d}.

We approximate the precession rate
\begin{equation}
\dot{\omega}_\textup{rel} \approx \frac{7.78}{1-e_d^2}\left(\frac{\Mstar}{M_\odot}\right) \left(\frac{a_d}{0.05\textup{ AU}}\right)^{-1} \left(\frac{P_d}{1\textup{ day}}\right)^{-1} \frac{\textup{degrees}}{\textup{century}} ,
\label{eq-rel}
\end{equation}
where the subscript $d$ refers to planet \D and each parameter is measured in the units of their respective normalization \citep{Jordan08}.
We evaluate Equation \ref{eq-rel} for 1,000 four-planet models described and find $\dot{\omega}_\textup{rel} = 3.45\pm^{0.05}_{0.03}$ degrees per century.
\citet{Baluev11} found that the effects of correlated noise results in a smaller estimate of $e_d$, though this does not change \wrel significantly.
The typical \wsec for the same models was $88\pm2$ degrees per century.

Ultimately, we wish to obtain a large sample of stable models but being careful not to neglect any mechanisms that could potentially stabilize a significant fraction of our models.
Therefore, relativistic precession will not be included in our final long-term integrations.


\section{Results for 3D Model + Long-Term Orbital Stability} 
\label{sec:3d}
The planet-planet interactions in GJ 876 are so strong that the physical planet masses must be used in the modeling process.
Next, we allow \RUNDMC to consider the full range of parameters associated with a three-dimensional (3-d) orbit for a four-planet model.
To provide a frame of reference, we set $\Omega_d=0\degree$.
For our set of initial conditions for \RUNDMC, we start with our posterior samples from \Mfour and perturb each inclination and ascending node value (excluding $\Omega_d$) by adding a number drawn randomly and uniformly between -1\degree\,and +1\degree.
\RUNDMC burned-in for about 20,000 generations until a long-term steady state was reached.
By modeling these extra variables, our parameter space grows from 32 dimensions (4 planets $\times$ 5 orbital elements $\{P, K, e, \omega, M\} + i_{sys} + $ 6 offsets + 5 jitters) to 38.

Additionally, we can obtain even more precise estimates by constraining the 3-d orbits of the planets from a direct analysis of the RV data and by demanding orbital stability.
In the end, we report 1,000 solutions where the planetary system is dynamically stable for at least $10^7$ years.
Our full procedure is described below and condensed into Table \ref{tbl-sets}.

We start by randomly drawing 10,000 posterior samples from our initial 3-D MCMC run (\setone) for our first set of stability tests.
The model parameter estimates for these samples are shown in Tables \ref{tbl-3d} and \ref{tbl-3d-sys}.
These systems are integrated for $10^5$ years and our conditions for stability are identical to those mentioned in \S \ref{sec:coplanar}.
Only about 11\% of our initial conditions passed our stability criterion; 48\% of our sample show planet \D falling into the central star, and 41\% of our sample show planet \E's semi-major axis suddenly changing by $\left|\triangle a/a\right| > 0.5$.
We visually inspected the results and determined which model parameters were most important for distinguishing between ``stable'' and ``unstable'' regimes of parameter space.
We find that the mutual inclination between planet pairs is the most sensitive parameter in regards to the system's orbital stability.
Henceforth, mutual inclinations between adjacent pairs of planets will be labeled as such: e.g. \imutdc for the mutual inclination between planets \D and \C (see Equation \ref{eq:mutinc}).
For planet \D, $60\degree<\imutdc<120\degree$ causes \D to undergo Kozai-like perturbations, pumping its eccentricity enough so that its periastron crosses the stellar surface (Figure \ref{fig:planetDstability}).
Figure \ref{fig:planetEstability} shows the joint parameter distribution for $\cos(\imutcb)$ and $\cos(\imutbe)$ based on \setone (contours).
We choose this parameterization over \imutcb and \imutbe to visually demonstrate how consistent our solutions are with a coplanar system.
For planet \E, all initial conditions with $\imutbe>13\degree$ ($\cos(\imutbe)<0.9781$) led to a sudden and large change in planet \E's semi-major axis.
Therefore, we perform a new set of \RUNDMC simulations in which we restrict the parameter space to exclude the unstable mutual inclinations found above, i.e. $60\degree<\imutdc<120\degree$, $\imutbe>13\degree$.
Although we see some large values of \imutcb correspond to an instability, we do not restrict the range of \imutcb for analysis in \settwo.
\RUNDMC is applied again on the same RV dataset, but now the resulting posterior sample (\settwo) is not as heavily diluted by wildly unstable models.

\begin{table*}
\centering
\caption{Estimates of the osculating orbital elements for all the known GJ 876 planets from self-consistent, 3-d dynamical fits. The top value in each cell are estimated directly from the radial velocities, and the bottom value imposes dynamical stability for $10^7$ years. }
\begin{tabular}{ccccc}
\hline
\hline
Parameter & Planet d & Planet c & Planet b & Planet e \\
\hline
\hline
\input{table-3d.tex}
\hline
\hline
\end{tabular}

\medskip
Estimates are computed using 15.9, 50, and 84.1 percentiles. Upper limits on $\Phi$ are based on a 99\% credible interval. For other details, refer to Section \ref{sec:3d}.
\label{tbl-3d}
\end{table*}

\begin{table*}
\centering
\caption{Systematic offset and jitter estimates based on a 3-d orbital model. }
\begin{tabular}{ccc}
\hline
Dataset & Offset [\ms] & Jitter [\ms] \\
\hline
\hline
\input{table-3d-sys.tex}
\hline
\hline
\end{tabular}

\medskip
Estimates are computed using 15.9, 50, and 84.1 percentiles. For other details, refer to Section \ref{sec:modelOfObs}.
\label{tbl-3d-sys}
\end{table*}

We repeat the above procedure one more time, integrating the \settwo for $10^6$ years.
We find the initial conditions are unstable, unless $0\degree<\imutcb<4\degree$ and $0\degree<\imutbe < 3\sqrt{16-\imutcb^2}\degree$.
We run  \RUNDMC once more while implementing these tighter constraints to obtain a new set of posterior samples for \setthree, which are subsequently integrated for $10^7$ years.
Parameter estimates of 1,000 stable solutions from \setthree are shown as the bottom value of each cell in Table \ref{tbl-3d}, as the solid green histogram in Figure \ref{fig:planetDstability}, and as green dots in Figure \ref{fig:planetEstability}.

Performing everything described above, we find the radial velocities alone place an upper limit of $\imutcb < 6.20\degree$ and $\imutbe < 28.5\degree$ based on a 99\% credible interval.
However if we expect the system to remain stable over the course of $10^7$ years, the system must be roughly coplanar: $\imutcb < 2.60\degree$ and $\imutbe < 7.87\degree$ based on a 99\% credible interval.



\section{Behavior of Resonant Angles}
\label{sec:angles}

\begin{figure*}
\begin{center}
\centerline{\includegraphics[scale=0.8]{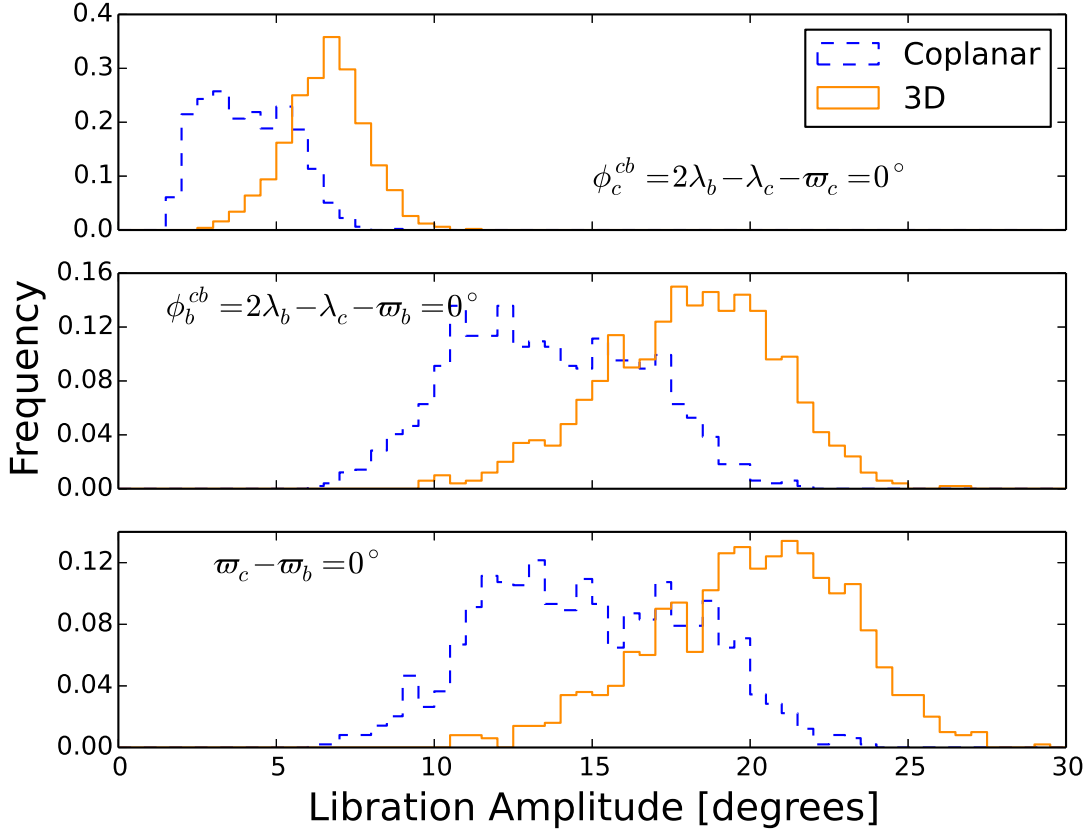}}
\caption{ Libration amplitude distributions for the three resonant angles of the \C and \B planet pair in a 2:1 mean-motion resonance. We compare libration amplitudes for a coplanar (dashed blue) and 3-d (solid orange) orbital model. The two angles associated with the 2:1 mean motion resonance (top and middle panels) are librating about 0\degree\,with low amplitude. The secular angle (bottom panel) also librates about 0\degree\,with low amplitude. }
\label{fig:angles_cb}
\end{center}
\end{figure*}

\subsection{Eccentricity Resonances}
With the final 3-d orbital solutions from \setthree of \S \ref{sec:3d}, we investigate the behavior of the critical angles associated with the mean-motion resonances relevant for this system.
For both our stable coplanar and 3-d orbital models, we compute the root-mean-square of the variability in each angle $\times \sqrt{2}$.
For a system undergoing small amplitude sinusoidal libration, this is an excellent approximation for the libration amplitude.

\begin{figure}
\begin{center}
\centerline{\includegraphics[width=0.9\columnwidth]{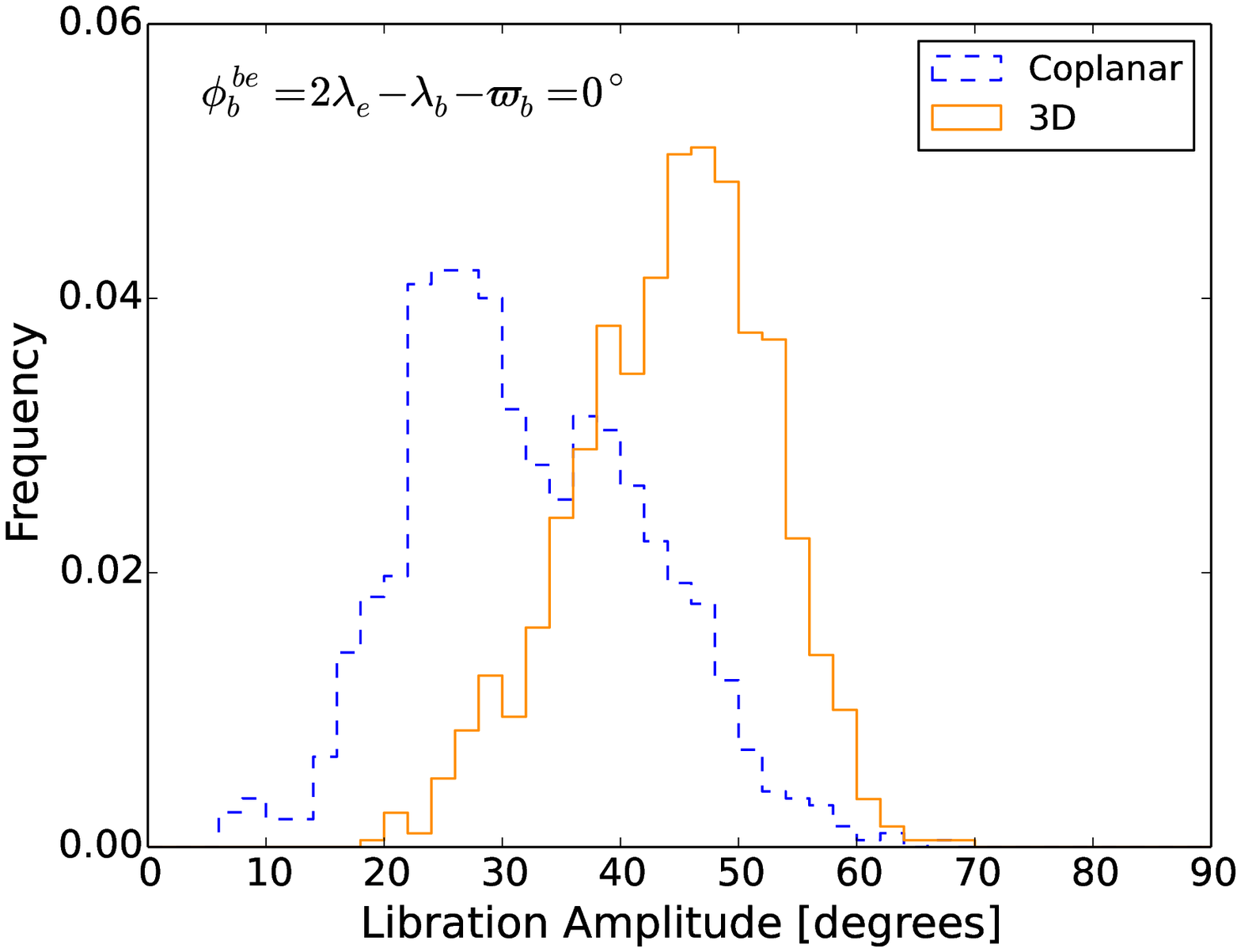}}
\caption{
Libration amplitude distributions for only librating angle of the \B and \E planet pair in a 2:1 mean-motion resonance.
We compare libration amplitudes for a coplanar (dashed blue) and 3-d (solid orange) orbital model.
For the vast majority of our sets of initial conditions, the other resonant argument and the secular angle are circulating. }
\label{fig:angles_be}
\end{center}
\end{figure}

\begin{figure}
\begin{center}
\centerline{\includegraphics[width=0.9\columnwidth]{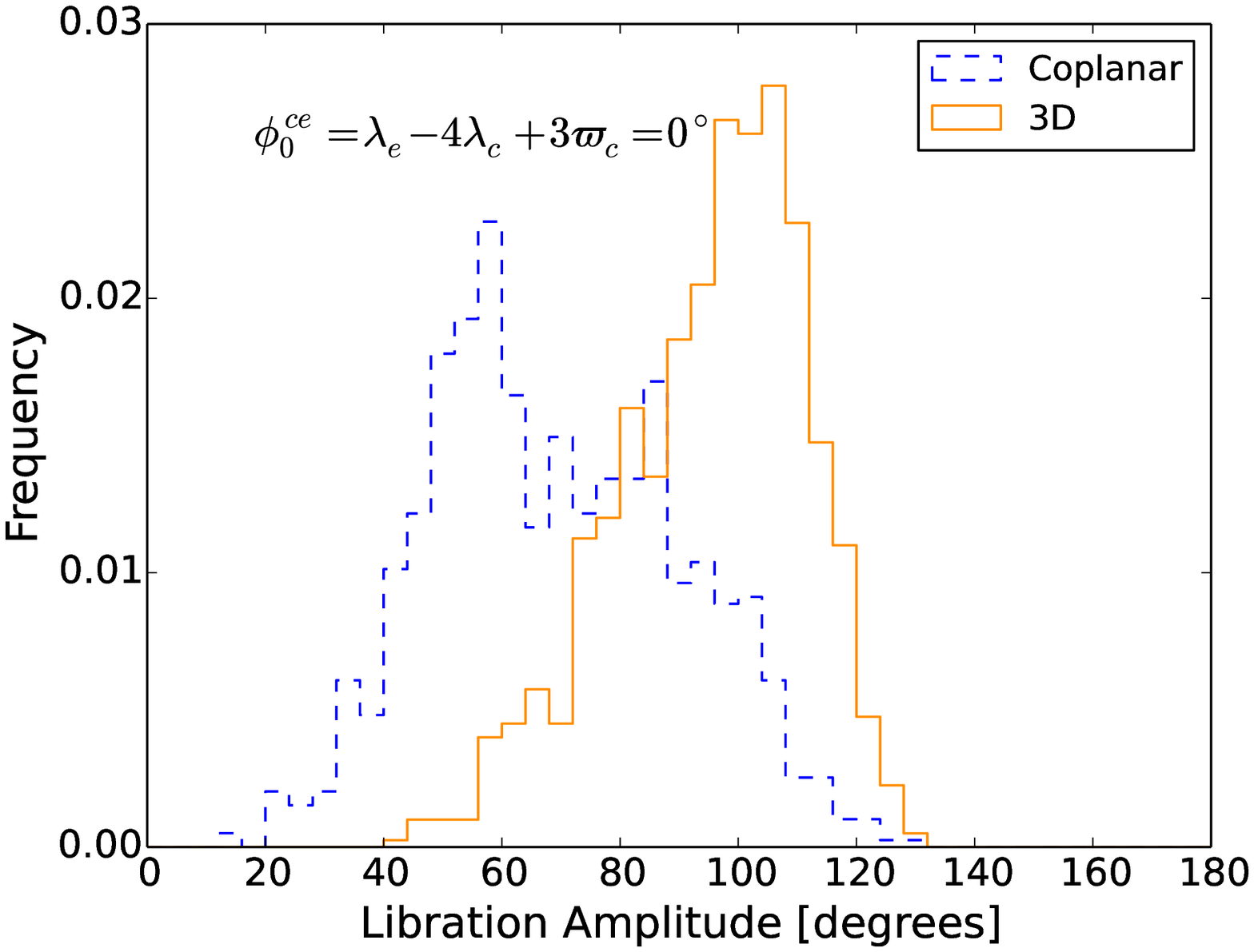}}
\caption{
Libration amplitude distribution for the only librating angle of the \C and \E planet pair relative to a 4:1 mean-motion resonance.
We compare libration amplitudes for a coplanar (dashed blue) and 3-d (solid orange) orbital model.
For the vast majority of our sets of initial conditions, the three other resonant arguments and the secular angle are circulating. }
\label{fig:angles_ce}
\end{center}
\end{figure}

For the \C and \B pair, the angles associated with the 2:1 MMR are the resonant angles, $\phi^{cb}_c = \lambda_c - 2\lambda_b + \varpi_c$ and $\phi^{cb}_b = \lambda_c - 2\lambda_b + \varpi_b$, and the secular angle $\varpi_c-\varpi_b$, where $\lambda$ and $\varpi$ are the mean longitude and longitude of pericenter respectively.
We find each of these angles is librating with low amplitude about 0\degree.
Figure \ref{fig:angles_cb} shows the distribution of the libration amplitude for angles associated with the \C and \B pair.
Similarly for the \B and \E pair, the angles associated with the 2:1 MMR are the resonant angles, $\phi^{be}_b = \lambda_b - 2\lambda_e + \varpi_b$ and $\phi^{be}_e = \lambda_b - 2\lambda_e + \varpi_e$, and the secular angle $\varpi_b-\varpi_e$.
We find $\phi^{be}_b$ librates about about 0\degree\,and the other two angles are circulating.
Figure \ref{fig:angles_be} shows the distribution of the libration amplitude for angles associated with the \B and \E pair.
For the \C and \E pair, the angles associated with the 4:1 MMR are the resonant angles, $\phi^{ce}_0 = \lambda_c - 4\lambda_e + 3\varpi_c$, $\phi^{ce}_1 = \lambda_c - 4\lambda_e + 2\varpi_c + \varpi_e$, $\phi^{ce}_2 = \lambda_c - 4\lambda_e + \varpi_c + 2\varpi_e$, and $\phi^{ce}_3 = \lambda_c - 4\lambda_e + 3\varpi_e$, and the secular angle $\varpi_c-\varpi_e$.
To distinguish these angles, the subscript refers to the multiplier in front of $\varpi_e$.
We find that four of the five angles are circulating.
$\phi^{ce}_0$ librates about 0\degree\,with low to medium amplitude.
Figure \ref{fig:angles_ce} shows the distribution of the libration amplitude for $\phi^{ce}_0$.
We report all of our libration amplitudes for coplanar and dynamically stable 3-d orbital models in Table \ref{tbl-angles}.

We find the secular angle $\omega_b-\omega_e$ is circulating, in contrast to previous studies  that reported libration about $180^\circ$.
This underscores the importance of performing self-consistent dynamical and statistical analyses when characterizing the evolution of interacting planetary systems.

\begin{table}
\centering
\caption{\small Libration Amplitudes of Resonant Angles. }
\begin{tabular}{ccc}
\hline
\hline
Angle & Amplitude, coplanar [\degree]  & Amplitude, 3D [\degree] \\
\hline
\hline
\input{table_angles.tex}
\hline
\hline
\end{tabular}

\medskip
See Section \ref{sec:angles} for details.
\label{tbl-angles}
\end{table}

The measured amplitude is interesting as it is neither so small as to imply strong dissipation nor so large as to suggest the absence of damping.
Given the importance of this result, we performed additional tests to verify that our algorithm accurately characterizes the libration amplitude of the Laplace angle for both regular and chaotic systems.
The methods and more detailed results of these test are presented in Appendix \ref{appendix}. 

\begin{figure}
\begin{center}
\centerline{\includegraphics[width=0.9\columnwidth]{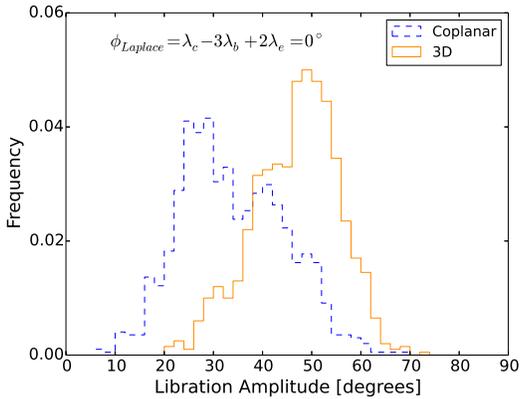}}
\caption{
Libration amplitude distributions for the Laplace argument of the \C, \B, and \E resonant trio.
We compare libration amplitudes for a coplanar (dashed blue) and 3-d (solid orange) orbital model. }
\label{fig:laplace}
\end{center}
\end{figure}

Upon the discovery of the outer-most planet, \citet{Rivera10} found that their best fit systems exhibit a 4:2:1 Laplace resonance.
The associated angle, $\Laplace = \lambda_c - 3\lambda_b + 2\lambda_e$, evolves \emph{chaotically}. 
Based on the Carnegie RVs alone, they found that \Laplace librates about $0\degree$ with an amplitude of $40\pm13\degree$ for a coplanar four-planet model.
Figure \ref{fig:laplace} shows our results for the posterior distribution for the libration amplitude for the Laplace argument for both the coplanar ($33\pm^{12.4}_{9.3}$\degree) and 3-d ($50.7\pm^{7.9}_{10}$\degree) cases.
The measured amplitude is interesting as it is neither so small as to imply strong dissipation nor so large as to suggest the absence of damping.
Given the importance of this result, we performed additional tests to verify that our algorithm accurately characterizes the libration amplitude of the Laplace angle for both regular and chaotic systems.
The methods and more detailed results of these test are presented in Appendix \ref{appendix}.

\subsection{Inclination Resonances}

Several studies have looked at inclination excitation and inclination resonance capture during planet migration phases \citep{ThommesLissauer03, Lee09, Libert09, Teyssandier14}.
The general conclusion in these studies is that inclination resonances can form during the formation phase but are likely to be short-lived.
A recent study by \citet{Barnes15} investigated the orbital evolution of MMR systems with mutually inclined orbits in the post-formation phase.
They find that their synthetic systems evolve chaotically but still retain dynamical stability and a MMR.
Drawing upon these conclusions, they model several real RV systems near a MMR to look for similar behavior.
They find that HD 73526 (a 2:1 MMR system) and HD 60532 (a 3:1 MMR system) could be evolving chaotically given a set of initial conditions listed in their Tables 9 and 10 respectively.
However, the inclination resonance arguments associated with these systems were all circulating, indicating that they were not in an inclination resonance.
To date, no exoplanet system has a detected inclination resonance.

We check for an inclination resonance in GJ 876 with our dynamically stable posterior samples.
First, we transform our set of initial conditions into the invariable plane\footnote{https://github.com/RoryBarnes/InvPlane} then integrate these for 1,000 years, logging the output every 10 days.
We compute the angles associated with the inclination-resonance for the inner pair $\phi^{cb}_{incl} = 4\lambda_b - 2\lambda_c - \Omega_c - \Omega_b$ and the outer pair $\phi^{be}_{incl} = 4\lambda_e - 2\lambda_b - \Omega_b - \Omega_e$.
Visually inspecting the orbital evolution over 100 years of these models, we find $\phi^{cb}_{incl}$ and $\phi^{be}_{incl}$ both circulate on timescales of $\sim$5 years.
While it is possible that $\phi_{incl}$ may ocassionally librate for a short period of time, this is sufficiently rare that we did not observe any such events over this baseline for all of our models.


\section{Summary and Discussion}
\label{sec:discussion}

We have meaningfully constrained the three-dimensional orbital architecture of the GJ 876 planetary system, based on 367 Doppler observations from several observing sites.

To verify the nature of the $\sim$124-day signal, we performed Bayesian model comparison of five different physical models, spanning three to five planets.
Since each evaluation of the likelihood requires an n-body integration for a strongly interacting planetary system like GJ 876, Bayesian model comparison becomes computationally costly.
Therefore, it was particularly important that we apply efficient and parallelizable algorithm.  
We refined and applied a modified importance sampling algorithm to compute the fully marginalized likelihood, or Bayesian evidence, starting from a posterior sample computed via MCMC methods \citep{Ford07,thesisPC,Weinberg13}.
The algorithm parallelizes readily and was implemented on GPUs using the \SWARMNG framework \citep{Dindar13}.  
While previous studies have computed Bayes factors using other algorithms (e.g. nested sampling \citep{Feroz09,Kipping13,Placek14}, geometric-path Monte Carlo \citep{Hou14}), these studies have assumed that the motion can be described as the linear superposition of Keplerian orbits, which is unsuitable for strongly interacting planetary systems such as GJ 876.
We believe this study to be the first example of rigorous Bayesian model comparison applied to strongly interacting planetary systems.
This algorithm is relatively easy to implement and worked well even for our high-dimensional ($\sim$30-40 parameter) models.

We determined that a four-planet model is most appropriate for the present data, based on a self-consistent Bayesian and n-body analysis (\S \ref{sec:noise} and \ref{sec:FML}).
When computing the evidence for our finite set of models, we can decisively choose a model with four signals with Bayes factors exceeding $10^{3}$.
We find a Bayes factor of $\sim$30 when comparing a four planet model (\Mfour) to a model with three planets plus a decoupled sinusoid, which could result from either a fourth planet in a circular orbit or a stellar activity signal (\Minstar).

On one hand, a Bayes factor of $\sim$30 is not large enough by itself to definitively choose the four planet model.  
On the other hand, we find no reason to suspect that stellar activity is masquerading as a planetary signal.
Looking at activity-sensitive indicators in publicly available HARPS spectra, we measure a stellar rotation period of $95\pm1$ days. 
This does not directly correspond with any of the planets' orbital periods.
As expected, the stellar rotation period and annual observing cycle lead to an alias with a frequency of $\sim$128 days.
In our coplanar model, the orbital period of \E is $124.5\pm1.3$, differing from the 128-day alias at slightly more than 2-$\sigma$ level.
If the signal were due solely to aliasing of the annual observing cycle with the stellar rotation period, then there should be an even stronger signal corresponding to the stellar rotation in the periodogram of the RV residuals (after subtracting the RV signal due to the inner-three planets).  
\citet{Rivera10} and \citet{Baluev11} looked for but did not find this effect, concluding that the signal is best explained by planet \E.

Through numerical integrations of systems where planet \E was treated as a test particle, \citet{Rivera10} explored the parameter space near the best-fit GJ 876 solution to the Carnegie RVs.  These tests suggested that the true system might be chaotic and was likely surrounded by regions of phase space with short-lived (unstable) orbits.
\citet{Marti13} studied the two reported 4-planet solutions to the radial velocity data using the MEGNO method (e.g. \citet{Cincotta00}) and found that these trajectories were both chaotic and in the Laplace resonance.
We show that the most probable Lyapunov time to be $\sim$10 years for both our coplanar and dynamically stable 3-d models (Appendix \ref{appendix:lyapunov}).
This timescale is consistent with the results of \citet{Batygin15} which were based on combining a simplified dynamical model and a point estimate for orbital parameters.

Strongly interacting planetary systems like GJ 876 could have unusually large transit timing variation amplitudes, enabling detailed characterization via the transit timing technique.
Upcoming missions such as TESS and PLATO would easily detect and confirm a similar system of transiting planets, but a detailed interpretation could be challenging given the anticipated observing timespans.
In particular, the perturbations from GJ 876 \E become evident only on multi-year timescales \citep{Libert13}.

While previous studies have assumed that the planets in GJ 876 follow coplanar orbits, we relax the assumption of coplanarity and allow for non-coplanar orbital configurations.  
We analyze the Doppler observations of GJ 876 and find that the three planets participating in the Laplace resonance must be nearly coplanar.
The 99\% credible upper limits on the mutual inclination constrained by just the RVs is impressive for both planet pairs: $\imutcb < 6.20\degree$ for the \C and \B pair and $\imutbe < 28.5\degree$ for the \B and \E pair.
Demanding orbital stability further restricts this range providing precise constraints on the mutual inclinations: $\imutcb < 2.60\degree$ and $\imutbe < 7.87\degree$.
A plot of the posterior samples suggest that the orbits of these planets are consistent with a coplanar system (Figure \ref{fig:planetEstability}).
Despite its rather unique orbital architecture, it seems that GJ 876 fits in with a population of M dwarf systems with several, coplanar planets \citep{BallardJohnson14}.

By performing the first self-consistent, Bayesian analysis of the four planets in GJ 876, we are able to accurately characterize the current dynamical state of all four planets and particularly the evolution of the resonant angle associated with the Laplace resonance. 
We measure the amplitude of variations to be $33\pm^{12.4}_{9.3}$ degrees for coplanar models or $50.5\pm^{7.9}_{10.0}$ degrees for fully three-dimensional models.

When measuring a positive definite quantity, observational uncertainties can bias measurements, particularly when using point estimates.
For example, the best-fit models of Doppler observations of a population of planets with nearly circular orbits will typically overestimate the planets' orbital eccentricities \citep{Zakamska11}, particularly when the Doppler amplitude is only a factor of a few greater than the measurement precision.
This motivated our Bayesian approach to characterizing the amplitude of variations of the Laplace angle.
Additionally, we performed tests of our algorithms using simulated planetary systems that confirm our algorithm accurately characterizes the behavior of the Laplace angle (Appendix \ref{appendix:methods}).
We conclude that formation theories for the GJ 876 system need to explain not only the resonant structure, but also the  chaotic evolution of the Laplace angle and its sizable amplitude.  

The near integer ratio in the planets' orbital periods are not likely a result of happenstance.
The probability of {\em in situ} formation yielding such a system is further reduced by the need to become trapped in a chaotic, but long-lived, multi-body resonance.  
Within the context of current planet formation models, the system most likely reached its current resonant configuration as the result of disk migration.
Migration through a smooth gas disk would be expected to result in strong damping of eccentricities, driving the system to a state where the resonant angle librates regularly with small amplitude.
Many of our simulations of a smooth migration with eccentricity damping lead to an amplitude of $\sim$1\degree, much smaller than we measure for the current system configuration.

While the observed large libration amplitude and chaotic evolution of resonant angles is contrary to the predictions of the simplest migration models, \citet{Batygin15} recently proposed that that the observed libration of the resonant angles can be explained purely by a turbulent migration.  
We propose an alternative formation mechanism based on a phase of smooth disk migration which terminates abruptly (Appendix \ref{appendix}).
In our simplistic migration models, turning off the eccentricity damping impulsively caused the libration amplitude for the Laplace angle to rise from $\sim1\degree$ to $\sim$tens of degrees.
Such a scenario could arise naturally as a result of rapid dispersal of inner disk via photoevaporation.  

Another potential possible formation scenario involves migration through a gas disk trapping the planets in resonance, followed by a phase of planet or planetessimal scattering.
\citep{Chatterjee15} showed that planetessimal scattering will naturally drive a planetary system initially in a 2:1 mean motion resonance farther apart, exciting eccentricities and perhaps even breaking the resonance.

It is possible for MMRs to arise from pure planet-planet scattering.
\citet{Raymond08b} found that MMRs could form through the ejection of one planet, based on simulations starting with three planets with drastically different masses.
One can distinguish between MMRs formed through scattering and migration mechanisms by the mass and orbital properties of the observed planets.
Scattering usually yields planet pairs in higher order resonances (second or greater), with larger semi-major axes ($>$1AU), and/or with a more massive outer companion.
The latter result could explain the \C and \B pair, but after considering \E and the observed Laplace resonance, planet-planet scattering alone becomes a less likely formation channel for the GJ 876 planets.
Indeed, \citet{Moeckel12} run a set of simulations modeling multi-planet systems during disk clearing phase and subsequent gas-free (purely Newtonian) phase and find that systems were much more likely to form resonant chains if scattering events did not occur.
In the presence of a planetessimal disk, already ``marginally stable'' systems can also form resonant chains \citep{Raymond09b}.

We encourage future studies to explore the predictions of these formation models for comparison to the GJ 876 system.
Continued long-term RV monitoring and/or astrometric observations from GAIA could continue to improve the dynamical constraints on this landmark system.

\section*{Acknowledgements}
We would like to thank our referee, Edward Thommes, for his constructive feedback that improved the manuscript.
We would like to thank Geoff Marcy and the entire of the California Planet Survey team for their long-term commitment to high-precision RVs for the GJ 876 system.
B.E.N. would like to thank Phil Gregory and Tom Loredo for useful suggestions regarding our methodology and presentation of our importance sampling algorithm.
Additionally, he would also like to thank Rory Barnes and Russell Deitrick for conversations regarding chaos and inclination resonances in multi-planet systems and Roman Baluev for useful feedback regarding the significance of correlated noise.

P.R. acknowledges support from NSF grant AST-1126413 and the Center for Exoplanets and Habitable Worlds.
M.J.P. gratefully acknowledges the NASA Origins of Solar Systems Program grant NNX13A124G.
E.B.F. and J.T.W. acknowledge NASA Keck PI Data Awards, administered by the NASA Exoplanet Science Institute, including awards 2007B N095Hr, 2010A N147Hr, 2011A\&B N141Hr, \& 2012A N129Hr.
This research was supported by NASA Origins of Solar Systems grant NNX09AB35G.
The authors acknowledge the University of Florida High Performance Computing Center and the Pennsylvania State Research Computing and Advanced Cyberinfrastructure Group for providing computational resources and support that have contributed to the results reported within this paper.
The Center for Exoplanets and Habitable Worlds is supported by the Pennsylvania State University, the Eberly College of Science, and the Pennsylvania Space Grant Consortium.
We extend special thanks to those of Hawai'ian ancestry on whose sacred mountain of Mauna Kea we are privileged to be guests.  
Without their generous hospitality, the Keck observations presented herein would not have been possible.

\bibliographystyle{aa}
\bibliography{references}

\clearpage

\appendix

\section{Fitting Synthetic Systems with Known Properties}
\label{appendix}

\citet{Batygin15} provide an elegant explanation of how such a chaotically evolving Laplace angle could have formed, and show that it can provide limits on the system's early formation:
\emph{smooth} migration can only form systems with relatively small libration amplitude, and \emph{regular} non-chaotic evolution;
\emph{stochastic} migration is required to form chaotically evolving systems with large libration amplitudes.

At face value, our results (\S \ref{sec:angles}) provide evidence for chaotic evolution of the Laplace argument in the GJ 876 system.
Given the importance of this result, we performed additional tests to determine whether low-libration amplitude is likely to be erroneously classified chaotic orbital evolution.
We simulated RV observations of a variety of \emph{synthetic} coplanar systems and then applied our \RUNDMC algorithm to verify that the recovered parameter estimates (in particular the libration amplitude of the Laplace angle) are consistent with their input values.


\subsection{Methodology: Creating Synthetic Planetary Systems and Datasets}
\label{appendix:methods}
Batygin (\emph{private communication}) provided orbital elements for the results of one of their \emph{stochastic migration} simulations.
We integrated this system forward, confirming that the evolution of the Laplace resonance angle, \Laplace, is indeed chaotic, and has a libration amplitude $\sim$80\degree.
This synthetic system will be referred to as \chaos henceforth (Figure \ref{fig:baty} left).

To complement this, we then performed a number of \emph{smooth migration} simulations in which the resultant simulations are regular and have low libration amplitudes. 
These migration simulations used the damped $a$ and $e$ method described in \citet{Lee02}, which was implemented in a modified version of \MERCURY \citep{Chambers99}.
We set up 3-planet systems so that each initial planetary period ratio was a little larger than 2:1 with their inner neighbor. 
Then, we applied the damped-migration model to the outer planet until it caught into a 2:1 resonance with the middle planet.
We continued the forced, damped migration until this outer resonant pair caught into a 4:2:1 resonance with the inner-most planet. 
Finally, we removed all forms of damping from the system and let them evolve for $\sim$$10^5$ years, ensuring that they were in an undamped equilibrium state.

We performed a number of these simulations and selected a system from these with final (i.e. after turning off orbital damping) libration amplitude of $\sim$20\degree.
Most simulations resulted in a Laplace argument with very small libration amplitude ($\sim$1\degree) during the damping phase and rose to several tens of degrees after the damping was removed.
This particular system had a libration amplitude significantly smaller than the best-fit from \S \ref{sec:angles}, allowing us to test whether our fitting procedure artificially increases the libration amplitude.
This synthetic system will be referred to as \synth henceforth (Figure \ref{fig:synth} left).

We generate an RV dataset for both \chaos and \synth using the same RV time series with their associated offsets and jitters described in \S \ref{sec:modelOfObs}.
We include the inner-most, non-resonant planet with the following orbital properties and coplanar with the migrated planets: $P_d=1.937821$ days, $m_d=2.21\times10^{-5} M_\odot$, $e_d=0.072$, $\omega_d=-137.09\degree$, and $M_d = 289.23\degree$.
For each synthetic measurement ($v_{obs}$), we compute the RV value at that time ($v_{mod}$), then add two noise terms: one drawn from a standard normal scaled by the measurement uncertainty $\sigma_{obs}$ and a similar term but scaled by the jitter value corresponding of that observation, i.e. $v_{obs} = v_{mod} + \mathcal{N}(0,1) \times \sigma_{obs} + \mathcal{N}(0,1) \times \jitter$.

These datasets were then used as simulated input observations for \RUNDMC, and a full differential evolution MCMC was performed on each synthetic data set (in the same manner as was performed on the real data, and described in \S \ref{sec:modelParameters} and \ref{sec:coplanar}).
We performed three realizations of the RV time series for both \chaos and \synth.


\subsection{Results for Synthetic Planetary Systems}
\label{appendix:results}

\begin{figure*}
\centering
\begin{tabular}{cc}
\includegraphics[width=1.0\columnwidth]{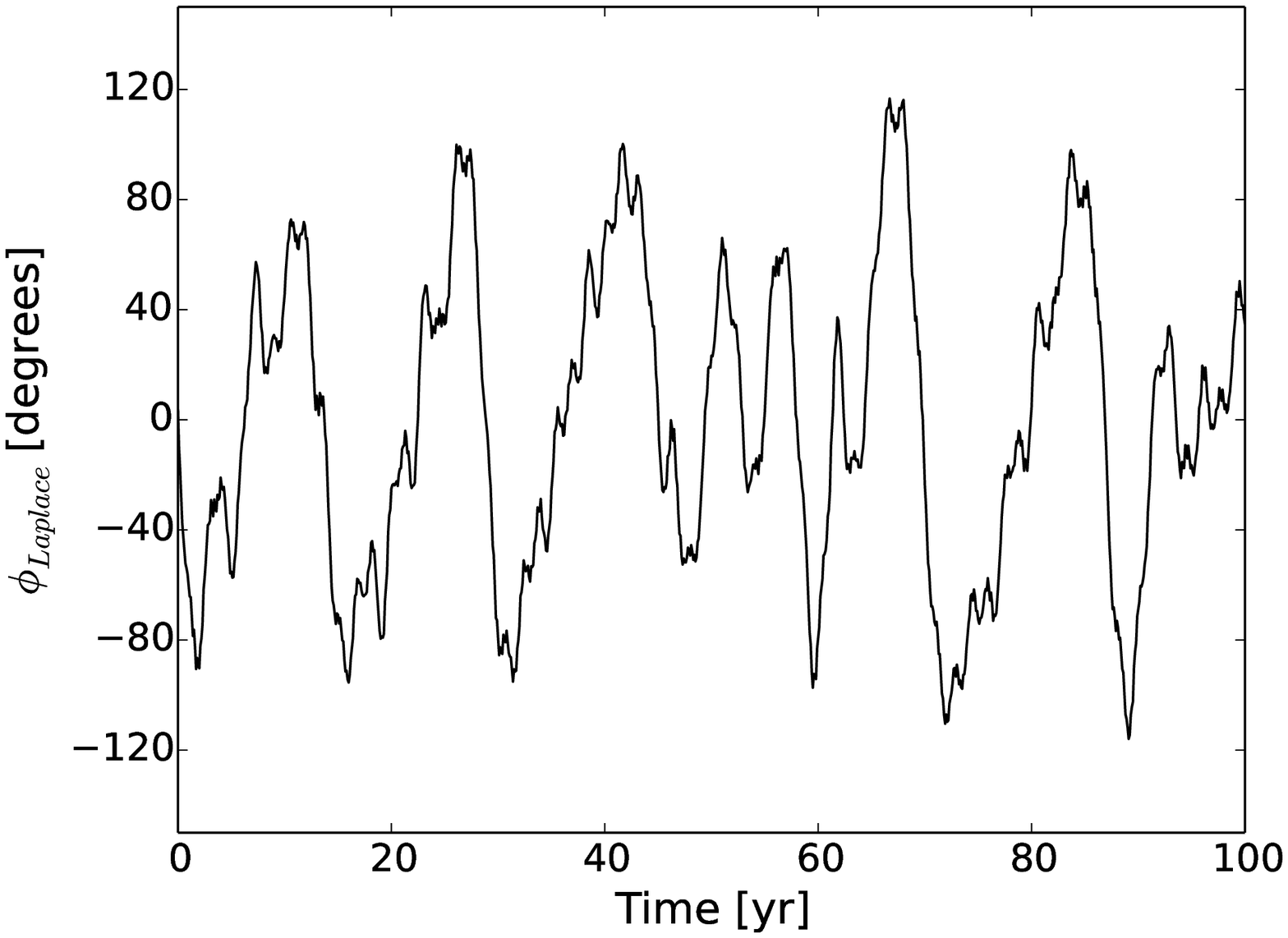} &
\includegraphics[width=1.0\columnwidth]{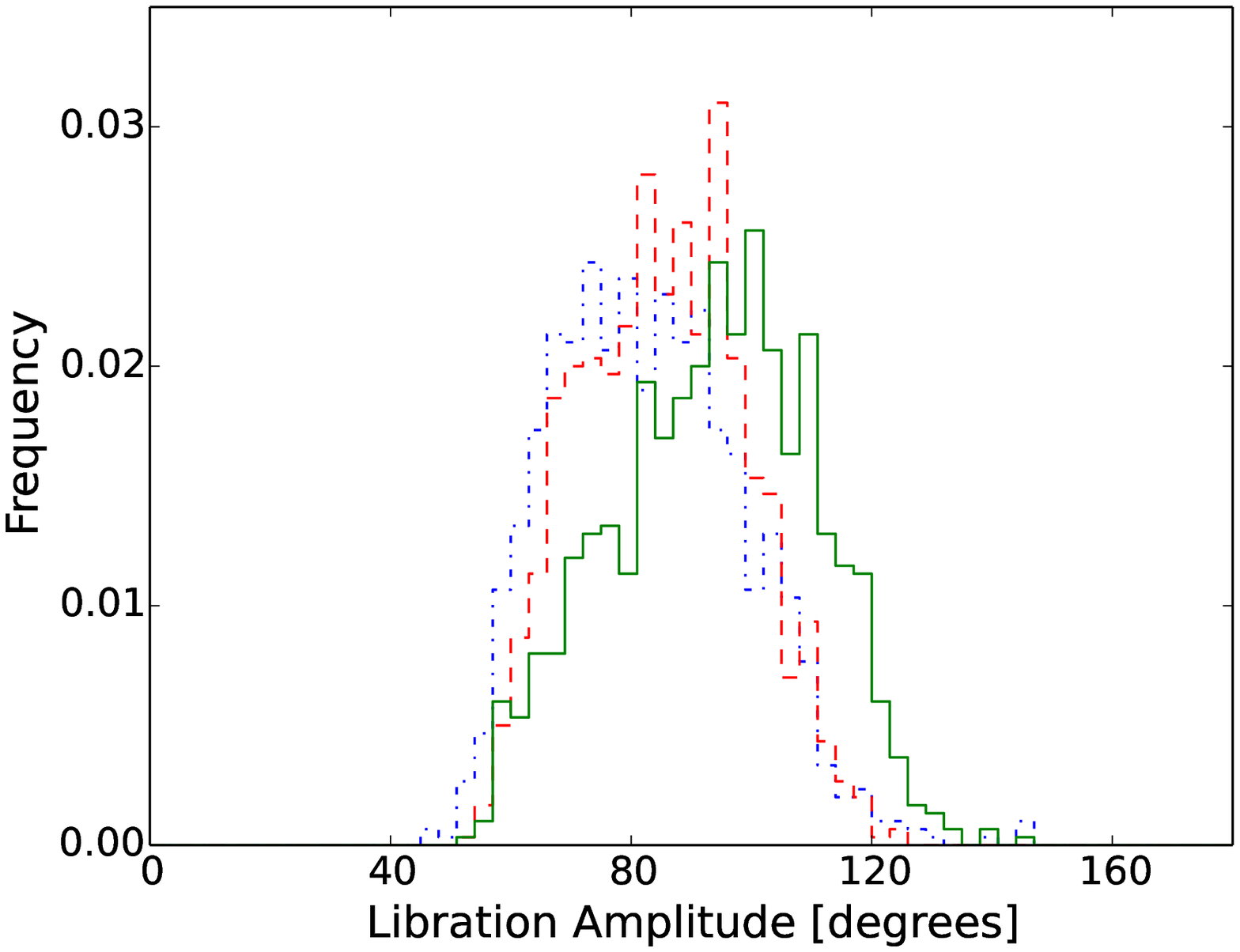} \\
\end{tabular}
\caption{
Radial velocity analysis of a synthetic system generated from a turbulent migration model, \chaos, from \citet{Batygin15}.
\textbf{Left}: The chaotic evolution of \Laplace over the course of 100 years for model \chaos.
\textbf{Right}: The distribution of the \Laplace libration amplitude based on three realizations (green solid, red dashed, blue dash-dotted) of the RV data generated from the synthetic system, \chaos.
}
\label{fig:baty}
\end{figure*}

\begin{figure*}
\centering
\begin{tabular}{cc}
\includegraphics[width=1.0\columnwidth]{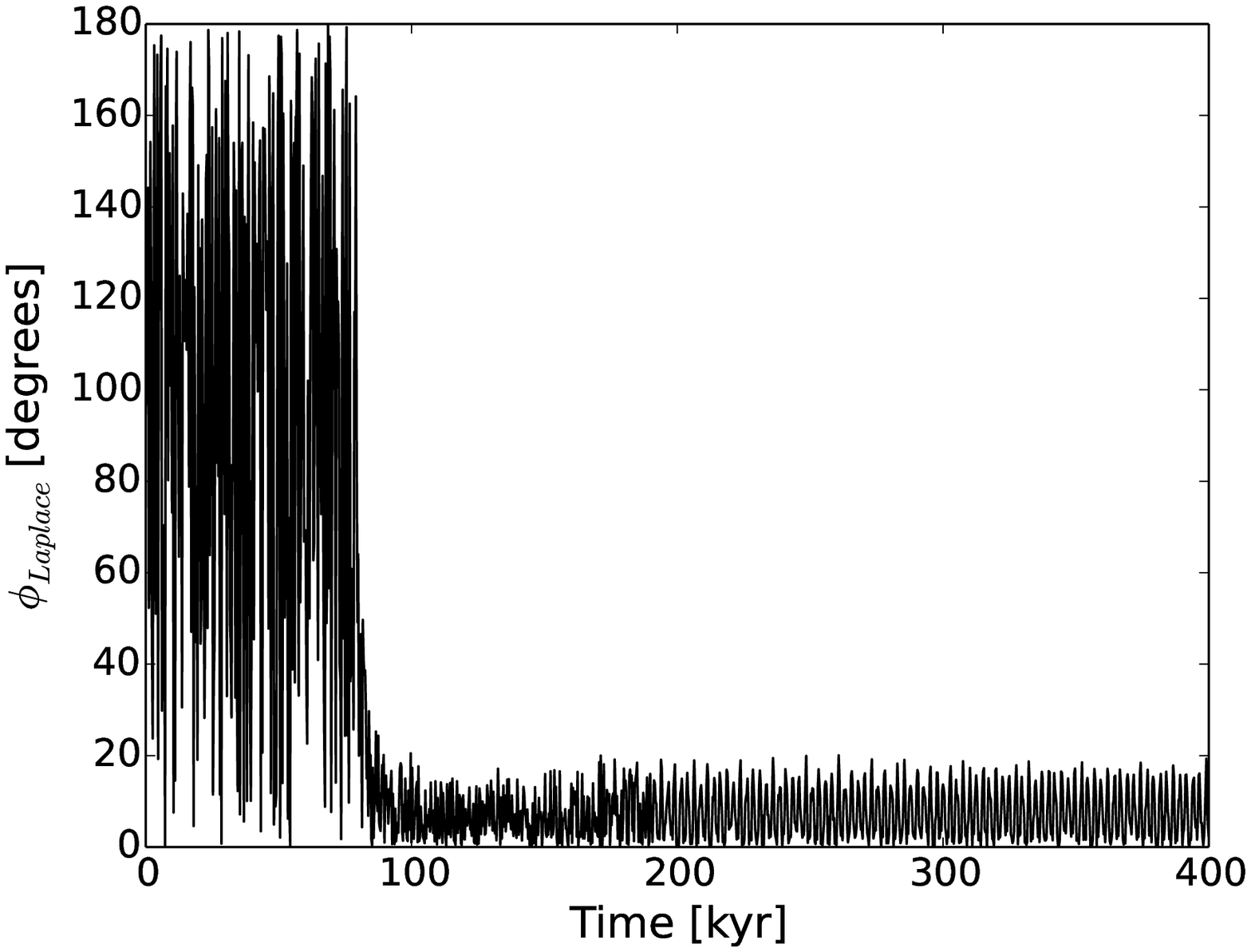} &
\includegraphics[width=1.0\columnwidth]{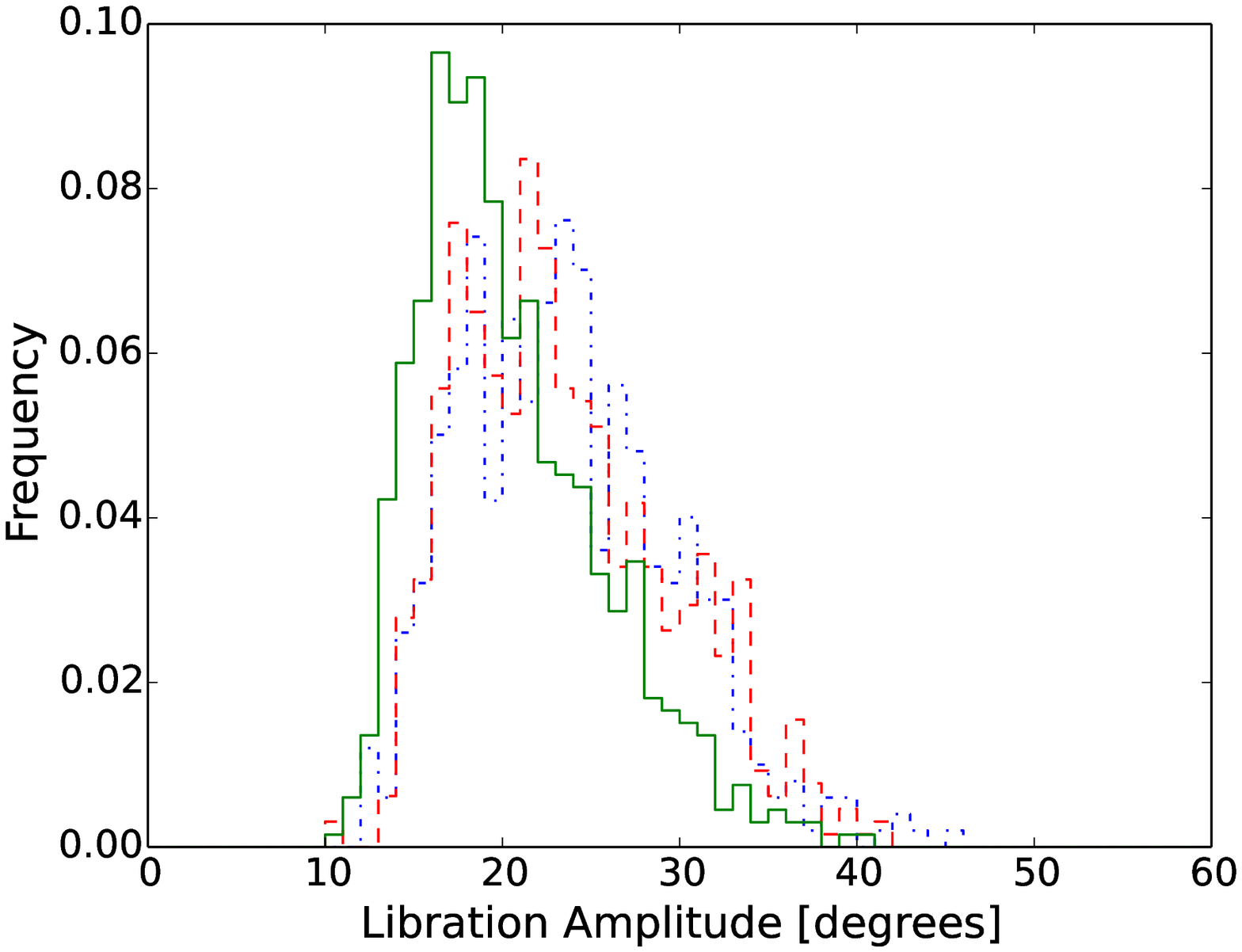} \\
\end{tabular}
\caption{
Radial velocity analysis of a synthetic system generated from a smooth migration model, \synth.
\textbf{Left}: The evolution of the absolute value of the \Laplace over the course of $4\times10^5$ years for model \synth.
\textbf{Right}: The distribution of the \Laplace libration amplitude based on three realizations (green solid, red dashed, blue dash-dotted) of the RV data generated from the synthetic system, \synth.
}
\label{fig:synth}
\end{figure*}

In right panel of Figure \ref{fig:baty}, we show the \Laplace libration amplitude distribution for the three realizations of the RV time series based on \chaos.
The chaotic nature of \chaos is shown in Figure \ref{fig:baty} left where the periodicity and peak-to-peak variation are not constant in time.
There is a rough variation of $\sim$160\degree\,corresponding to an amplitude of $\sim$80\degree.
For each realization, we find libration amplitudes of $97\pm^{14}_{19}$ (blue dash-dotted), $96\pm^{16}_{21}$ (red dashed), and $91\pm^{15}_{17}$ (green solid) degrees based on 1,000 solutions each.
Only a few solutions from each of these realizations were unstable over the short integration baseline (1 kyr).
These estimates are qualitatively consistent with the input amplitude.

In right panel of Figure \ref{fig:synth}, we show the \Laplace libration amplitude distribution for the three realizations of the RV time series based on \chaos.
A significant fraction of our posterior sample for each realization was not dynamically stable.
Specifically, only 548, 496, and 673 systems remained after the 1 kyr integration.
We find that the RV analysis of the synthetic data overestimated the outer-most planet's mass by $\sim$20\%, which could be reason for such a large fraction of unstable systems.
We find libration amplitudes of $25.6 \pm^{4.7}_{5.3}$ (blue dash-dotted), $22.9\pm^{6.8}_{5.5}$, (red dashed) and $19.5\pm^{5.6}_{3.7}$ (green solid) degrees based on the remaining short-term stable systems.
Again, these values are qualitatively consistent with the input amplitude of $\sim$20\degree.

It is clear from these results that our \RUNDMC algorithm does \emph{not} significantly bias the recovered libration amplitudes for the 4:2:1 resonance of the outer three planets in the GJ 876 system.
Thus, we believe that our best-fit libration amplitude for the real data (\S \ref{sec:angles}) is accurate and the GJ 876 system is in a chaotically librating state, with a high-amplitude libration.

\subsection{Estimation of Lyapunov times for GJ 876 Orbital Solutions}
\label{appendix:lyapunov}

\citet{Batygin15} suggests that the GJ 876 system is chaotic, with a characteristic timescale (or Lyapunov time, defined below) for the chaos of roughly 14 years.
This theoretical study ignored the innermost planet and treated the system as coplanar.
An estimate of the Lyapunov time for the 4-planet, coplanar radial velocity orbital solution of \citet{Rivera10} was consistent with the analytic estimate.
\citet{Barnes15} suggests that mutually inclined systems in or near a MMR can also exhibit chaotic evolution.

Here we study the nature of three sets of solutions: first, the 1,000 long-term stable, 3-d orbital models described in \S \ref{sec:3d}; 1,000 coplanar models described in \S \ref{sec:coplanar}; and 1,000 coplanar models based on the \synth.
We determine whether or not the orbits are chaotic by evolving simultaneously the standard gravitational equations of motion and the variational equations of motion, which yields an estimate of the Lyapunov time of a trajectory (e.g. \citet{Lichtenberg92}).
The variational equations govern the behavior of small perturbations to an orbit and therefore can be used to study how perturbations evolve in time.
For chaotic orbits, small perturbations of length $D$ grow exponentially as $D \sim e^{t/\tau}$ with a characteristic time $\tau$, which in the limit as $t \rightarrow \infty$ is defined as the (minimum) Lyapunov time.
Our finite time integrations are used to estimate this Lyapunov time as $t_{final}/\log{[(D(t_{final})]}$, where $t_{final}$ is the total integration time, $D(t=0)=1$, and $D(t_{final})$ is the total length of the ``perturbation" at the end of the integration.
Note that $D$ in principle can become very large, but since the variational equations are linear in the components of $D$, the absolute length of $D$ need not be small for the variational equations to apply.
If an orbit is regular, the reported Lyapunov time will be comparable to the integration time, though integrations cannot prove an orbit is regular.
These integrations must therefore be carried out for long enough such that the chaotic and regular orbits have markedly different reported Lyapunov times. 

We employed a Wisdom-Holman mapping in canonical astrocentric coordinates to integrate both the equations of motion and the variational equations \citep{Wisdom91}.
A third-order symplectic corrector was implemented to improve the accuracy of these integrations \citep{Wisdom96,Chambers99,Wisdom06}.
We used a simple prescription for the effects of general relativity which leads to precession of the orbits with the correct timescale \citep{Milani83}.
As we will show, the Lyapunov times of the orbits were significantly shorter than the timescale of precession due to general relativity, and so this approximate version of general relativity is  sufficient for a good estimate of the Lyapunov times for the orbits studied. 

The Wisdom-Holman integrator can be used with a time step as large as a tenth or twentieth of the shortest orbital timescale in the system \citep{Wisdom92,Rauch99}.
For GJ 876, this corresponds to the time needed to resolve the pericenter passage of the innermost planet.
We estimate this using the orbital period of the innermost planet as if its semimajor axis was equal to the pericenter distance, $P_{eff} \sim P_{orb} (1-e)^{3/2}= 1.6$ days, where $e \approx 0.1$ and $P_{orb}\approx 2$ days. 
We used a time step of either 0.14 or 0.014 days when integrating the set of 3-d solutions, and, as discussed below, these yield similar estimates of the Lyapunov times.
For both coplanar set of solutions, we used a time step of 0.14 days.
The maximum fractional energy error $\Delta E/E$ was  $\sim10^{-5}$ for the 3-d set of orbits with a time step of 0.14 days, $\sim 10^{-10}$ for the same set with a time step of 0.014 days, and $\sim 10^{-7}$ for the coplanar set of orbits.
These integrations lasted $10^4$ years.

\begin{figure}[h]\label{fig:distribution}
\begin{center}
\includegraphics[width=0.9\columnwidth]{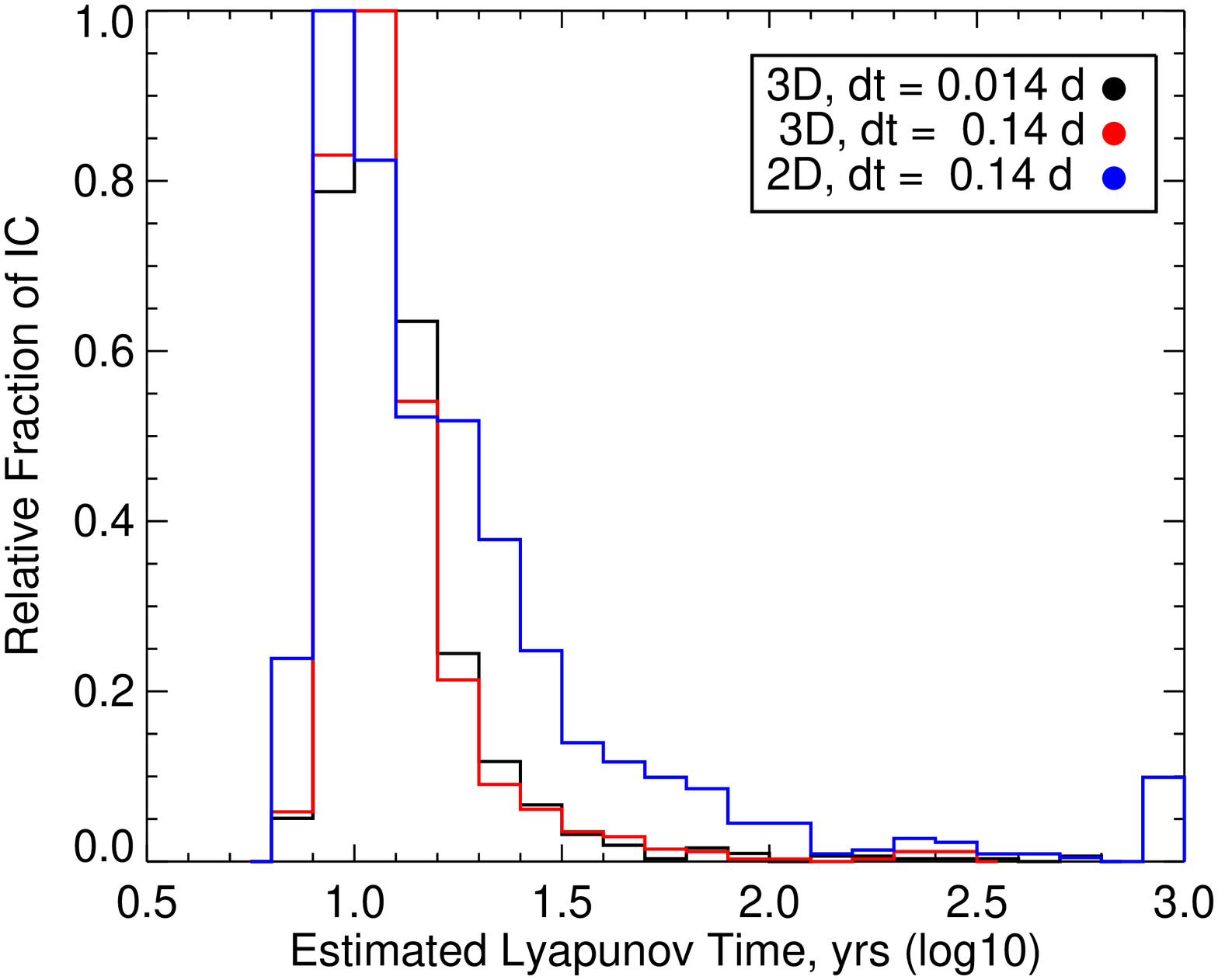}
\caption{The distribution of estimated Lyapunov times for the GJ-876 system.
These estimates are the result of integrations of the set of $10^3$ 3-d orbital solutions fit to the radial velocity data, employing a time step of 0.14 days (red) and with a time step of 0.014 days (black), and for the 1,000 solutions fit to the data assuming coplanar orbits, which employed a time step of 0.14 days (blue).}
\end{center}
\end{figure}

In Figure \ref{fig:distribution}, we show the distribution of Lyapunov times for the 3-d set of solutions resulting from the integrations employing both timesteps, and the distribution of Lyapunov times for the coplanar set of solutions.
These integrations all agree that the motion of these orbits is chaotic with a Lyapunov time of roughly 10 years, consistent with the analytic estimate of \citet{Batygin15}.
The small peak at $10^3$ years in the distribution of Lyapunov times for the coplanar orbital solutions to the data corresponds to orbits which might be regular.
However, the vast majority of coplanar orbits studied were chaotic.
The close agreement between Lyapunov times estimated for the 3-d and coplanar sets is also consistent with the analysis of \citet{Batygin15} and \citet{Marti13} in that the chaotic motion is captured by a coplanar approximation of the true system. 

During the integrations with a time step of 0.14 days, $\sim$10 orbits in each set (3-d vs. coplanar) were flagged as unstable.
With the smaller time step of 0.014 days, no orbit in the set of 3-d orbits was flagged as unstable.
This suggests that the smaller time step might be necessary for studying the long-term stability of these orbits using a Wisdom-Holman integrator.
However, since the two time steps agree on the distribution of Lyapunov times, we believe these results are robust.
Lastly, we do not show the results of the coplanar set of solutions fit to synthetic data, since our integrations suggested that nearly all of these orbits where not only chaotic but also showed instability on short ($10^4$ year) timescales.
We defer any further analysis of the stability of these synthetic solutions to future work.

\clearpage

\label{lastpage}
\end{document}

%% file: table_rvs.tex
3301.808032 & -33.89 & 2.06 \\
3301.816875 & -36.49 & 2.37 \\
3301.822986 & -28.19 & 2.60 \\
3301.871412 & -39.54 & 1.99 \\
3302.722523 & -91.03 & 1.91 \\
3302.728866 & -89.20 & 2.02 \\
3302.735567 & -81.93 & 1.92 \\

%% file: table_bf.tex
$\Mfive$ & $\textup{\Mfour + injected fifth}$ & $743.8\pm^{11.8}_{9.5}$ & $-1255.159$ & \multirow{2}{*}{$\sim10^{-3}$} \\ 
$\Mfour$ & $\textup{all four planets}$ & $768.3\pm^{9.7}_{8.0}$ & $-1248.139$ &  \multirow{2}{*}{$\sim30$} \\
$\Minstar$ & \textup{\Min + $\sim$120 day sinusoid} & $779.7\pm^{8.7}_{7.5}$ & $-1251.671$ &  \multirow{2}{*}{$\sim10^{9}$} \\
$\Min$ & $\textup{3 inner-most planets}$  & $877.4\pm^{8.3}_{6.9}$ & $-1273.519$ & \multirow{2}{*}{$\sim10^{31}$} \\
$\Mout$ & $\textup{3 outer-most planets}$ & $1045.7\pm^{8.7}_{7.9}$ & $-1345.970$ & \\

%% file: table-cop.tex
{$P$ [d]} & $1.937848 \pm^{0.000024}_{0.000023}$ & $30.0784 \pm^{0.0075}_{0.0071}$ & $61.087 \pm^{0.011}_{0.012}$ & $124.50 \pm 1.33$ \\
{$K$ [m\,s$^{-1}$]} & $6.01 \pm 0.30$ & $87.93 \pm 0.36$ & $213.94 \pm^{0.37}_{0.38}$ & $3.31 \pm^{0.38}_{0.42}$ \\
{$m$ $\frac{0.37 \textup{M}_\odot}{\textup{M}_\star}$ [M$_\oplus$]} & $7.49 \pm^{0.43}_{0.41}$ & $266.2 \pm^{4.8}_{4.3}$ & $845.3 \pm^{16.6}_{15.0}$ & $16.58 \pm^{1.96}_{2.11}$ \\
{$a$ [AU]} & $0.02183911 \pm^{0.00000019}_{0.00000018}$ & $0.135989 \pm^{0.000023}_{0.000021}$ & $0.218585 \pm 0.000026$ & $0.3514 \pm 0.0025$ \\
{$e$} & $0.119 \pm^{0.050}_{0.052}$ & $0.2531 \pm 0.0031$ & $0.0368 \pm^{0.0019}_{0.0025}$ & $0.031 \pm^{0.030}_{0.025}$ \\
{$e\sin\omega$} & $-0.044 \pm^{0.052}_{0.059}$ & $0.2276 \pm 0.0045$ & $0.0345 \pm^{0.0020}_{0.0032}$ & $0.008 \pm^{0.013}_{0.011}$ \\
{$e\cos\omega$} & $-0.097 \pm^{0.050}_{0.046}$ & $-0.1108 \pm^{0.0049}_{0.0042}$ & $-0.0130 \pm^{0.0022}_{0.0020}$ & $-0.013 \pm^{0.021}_{0.037}$ \\
{$i$ [degrees]} & \multicolumn{4}{c}{$53.19 \pm^{1.39}_{1.43}$} \\
{$\omega$ [degrees]} & $-140.90 \pm^{286.32}_{23.85}$ & $115.96 \pm^{1.21}_{1.35}$ & $110.80 \pm^{3.86}_{3.91}$ & $129.56 \pm^{36.31}_{185.78}$ \\
{$M$ [degrees]} & $301.87 \pm^{24.82}_{286.24}$ & $-220.68 \pm^{1.38}_{1.35}$ & $-285.38 \pm^{4.15}_{4.11}$ & $-175.89 \pm^{190.80}_{41.44}$ \\
{$\omega+M$ [degrees]} & $161.71 \pm^{6.43}_{6.71}$ & $-104.74 \pm^{0.63}_{0.64}$ & $-174.62 \pm^{0.56}_{0.45}$ & $-45.05 \pm^{9.73}_{10.97}$ \\

%% file: table-cop-sys.tex
{HIRES Carnegie (pre-upgrade)} & $50.48 \pm^{0.34}_{0.33}$ & \multirow{2}{*}{$2.38 \pm^{0.25}_{0.23}$} \\
{HIRES Carnegie (post-upgrade)} & $52.20 \pm^{0.56}_{0.57}$ &  \\
{HIRES California (post-upgrade)} & $8.18 \pm 0.94$ & $6.82 \pm^{0.74}_{0.63}$ \\
{ELODIE} & $-1903.81 \pm^{4.48}_{4.76}$ & $18.56 \pm^{4.73}_{3.94}$ \\
{CORALIE} & $-1864.40 \pm^{3.76}_{3.72}$ & $20.36 \pm^{3.30}_{2.84}$ \\
{HARPS} & $-1337.64 \pm^{0.44}_{0.43}$ & $1.63 \pm^{0.26}_{0.24}$ \\

%% file: table-3d.tex
\multirow{2}{*}{$P$ [d]} & $1.937891 \pm^{0.000037}_{0.000034}$ & $30.0758 \pm^{0.0084}_{0.0078}$ & $61.094 \pm 0.013$ & $123.50 \pm^{1.65}_{1.33}$ \\
 & $1.937870 \pm^{0.000025}_{0.000028}$ & $30.0766 \pm^{0.0073}_{0.0065}$ & $61.087 \pm^{0.011}_{0.012}$ & $124.72 \pm^{1.26}_{1.41}$ \\ \hline
\multirow{2}{*}{$K$ [m\,s$^{-1}$]} & $6.13 \pm^{0.39}_{0.35}$ & $88.74 \pm^{0.56}_{0.62}$ & $213.63 \pm^{0.50}_{0.48}$ & $3.18 \pm^{0.51}_{0.40}$ \\
 & $6.11 \pm^{0.35}_{0.40}$ & $88.33 \pm^{0.47}_{0.44}$ & $213.71 \pm^{0.43}_{0.41}$ & $3.44 \pm^{0.42}_{0.41}$ \\ \hline
\multirow{2}{*}{$m$ $\frac{0.37 \textup{M}_\odot}{\textup{M}_\star}$ [M$_\oplus$]} & $7.03 \pm^{3.55}_{0.90}$ & $274.2 \pm^{10.1}_{10.0}$ & $850.5 \pm^{16.6}_{16.0}$ & $15.45 \pm^{2.99}_{2.32}$ \\
 & $6.90 \pm^{3.40}_{0.81}$ & $267.9 \pm^{6.1}_{5.9}$ & $848.5 \pm^{16.4}_{17.1}$ & $17.16 \pm^{2.18}_{1.96}$ \\ \hline
\multirow{2}{*}{$a$ [AU]} & $0.02183945 \pm^{0.00000034}_{0.00000026}$ & $0.135984 \pm^{0.000026}_{0.000023}$ & $0.218609 \pm^{0.000034}_{0.000031}$ & $0.3495 \pm^{0.0031}_{0.0025}$ \\
 & $0.02183930 \pm^{0.00000022}_{0.00000020}$ & $0.135985 \pm^{0.000022}_{0.000019}$ & $0.218589 \pm^{0.000026}_{0.000030}$ & $0.3518 \pm^{0.0024}_{0.0026}$ \\ \hline
\multirow{2}{*}{$e$} & $0.113 \pm^{0.053}_{0.050}$ & $0.2532 \pm^{0.0032}_{0.0031}$ & $0.0371 \pm^{0.0019}_{0.0029}$ & $0.046 \pm^{0.041}_{0.030}$ \\
 & $0.108 \pm^{0.050}_{0.047}$ & $0.2539 \pm^{0.0032}_{0.0034}$ & $0.0365 \pm^{0.0019}_{0.0030}$ & $0.031 \pm^{0.027}_{0.026}$ \\ \hline 
 \multirow{2}{*}{$e\sin\omega$} & $-0.038 \pm^{0.049}_{0.058}$ & $0.2208 \pm^{0.0087}_{0.0072}$ & $0.0345 \pm^{0.0021}_{0.0034}$ & $0.019 \pm^{0.027}_{0.028}$ \\
 & $-0.034 \pm^{0.049}_{0.055}$ & $0.2259 \pm^{0.0052}_{0.0050}$ & $0.0340 \pm^{0.0022}_{0.0038}$ & $0.004 \pm^{0.014}_{0.013}$ \\ \hline
 \multirow{2}{*}{$e\cos\omega$} & $-0.090 \pm^{0.049}_{0.051}$ & $-0.1236 \pm^{0.0135}_{0.0101}$ & $-0.0137 \pm^{0.0024}_{0.0021}$ & $-0.025 \pm^{0.033}_{0.040}$ \\
 & $-0.088 \pm^{0.049}_{0.046}$ & $-0.1160 \pm^{0.0073}_{0.0065}$ & $-0.0137 \pm^{0.0022}_{0.0021}$ & $-0.006 \pm^{0.020}_{0.038}$ \\ \hline
\multirow{2}{*}{$i$ [degrees]} & $87.44 \pm^{41.97}_{40.94}$ & $51.63 \pm^{2.64}_{2.32}$ & $52.61 \pm^{1.44}_{1.36}$ & $55.51 \pm^{7.34}_{5.89}$ \\
 & $88.26 \pm^{38.90}_{40.80}$ & $53.06 \pm^{1.77}_{1.72}$ & $52.82 \pm^{1.54}_{1.44}$ & $53.29 \pm^{3.16}_{3.17}$ \\ \hline
\multirow{2}{*}{$\omega$ [degrees]} & $-139.25 \pm^{288.00}_{26.97}$ & $119.26 \pm^{2.73}_{3.58}$ & $111.81 \pm^{4.57}_{4.13}$ & $122.34 \pm^{33.00}_{207.55}$ \\
 & ${162.52}\pm^{6.67}_{6.74}$ & $117.12 \pm^{1.70}_{1.74}$ & $112.27 \pm^{4.51}_{4.13}$ & $-54.2\pm^{23.7}_{24.0}$ \\ \hline
\multirow{2}{*}{$M$ [degrees]} & $301.24 \pm^{27.44}_{289.02}$ & $-223.46 \pm^{3.17}_{2.78}$ & $-286.01 \pm^{4.30}_{5.16}$ & $-184.28 \pm^{231.29}_{38.26}$ \\
 & $300.58 \pm^{25.88}_{294.52}$ & $-221.76 \pm^{2.00}_{1.72}$ & $-286.82 \pm^{4.24}_{4.85}$ & $-92.97 \pm^{154.83}_{119.00}$ \\ \hline
 \multirow{2}{*}{$\omega+M$ [degrees]} & $162.52 \pm^{6.67}_{6.74}$ & $-104.33 \pm^{0.73}_{0.71}$ & $-174.26 \pm^{0.57}_{0.76}$ & $-54.24 \pm^{23.68}_{23.98}$ \\
 &  $162.28 \pm^{6.77}_{7.28}$ & $-104.60 \pm^{0.57}_{0.61}$ & $-174.64 \pm^{0.46}_{0.45}$ & $-42.46 \pm^{11.62}_{8.91}$ \\ \hline
\multirow{2}{*}{$\triangle\Omega$ [degrees]} & & \multirow{2}{*}{$\triangle\Omega_{cb}$ = } $-2.72\pm^{2.02}_{1.73}$ & & \multirow{2}{*}{$\triangle\Omega_{be}$ = } $-5.01\pm^{15.17}_{10.91}$  \\
& & \,\,\,\,\,\,\,\,\,\,\,\,\,\,\,\,\,\,\,\,\,\,\,\,\,$-1.29 \pm^{0.96}_{0.89}$ & & \,\,\,\,\,\,\,\,\,\,\,\,\,\,\,\,\,\,\,\,\,\,\,\,\,\,\,$1.29 \pm^{4.00}_{4.58}$ \\ \hline
\multirow{2}{*}{$\Phi$ [degrees]} & & \multirow{2}{*}{$\imutcb <$} $6.20$ & & \multirow{2}{*}{$\imutbe <$} $28.5$  \\
& & \,\,\,\,\,\,\,\,\,\,\,\,\,\,\,\,\,\,\,$2.60$ & & \,\,\,\,\,\,\,\,\,\,\,\,\,\,\,\,\,\,\,\,\,$7.87$ \\

%% file: table-3d-sys.tex
\multirow{2}{*}{HIRES Carnegie (pre-upgrade)} & $50.44 \pm 0.34$ &  \\
 & $50.49 \pm 0.31$ & $2.40 \pm^{0.25}_{0.24}$ \\ 
 \cline{1-2} 
\multirow{2}{*}{HIRES Carnegie (post-upgrade)} & $52.04 \pm^{0.56}_{0.55}$ & $2.36 \pm^{0.24}_{0.23}$ \\
 & $52.12 \pm^{0.65}_{0.56}$ & \\ \hline
\multirow{2}{*}{HIRES California (post-upgrade)} & $8.02 \pm 0.90$ & $6.76 \pm^{0.75}_{0.64}$ \\
 & $8.05 \pm^{0.95}_{0.92}$ & $6.93 \pm^{0.74}_{0.64}$ \\ \hline
\multirow{2}{*}{ELODIE} & $-1904.25 \pm^{4.58}_{4.81}$ & $18.97 \pm^{4.70}_{4.04}$ \\
 & $-1903.83 \pm^{4.65}_{4.88}$ & $18.69 \pm^{5.18}_{3.81}$ \\ \hline
\multirow{2}{*}{CORALIE} & $-1864.32 \pm^{3.53}_{3.60}$ & $19.96 \pm^{3.03}_{2.78}$ \\
 & $-1864.48 \pm^{3.65}_{3.52}$ & $19.94 \pm^{3.19}_{2.66}$ \\ \hline
\multirow{2}{*}{HARPS} & $-1337.68 \pm^{0.50}_{0.47}$ & $1.52 \pm^{0.27}_{0.24}$ \\
 & $-1337.58 \pm^{0.41}_{0.42}$ & $1.57 \pm^{0.25}_{0.22}$ \\

%% file: table_angles.tex
$\phi^{cb}_c$ & $4.0\pm^{1.6}_{1.4}$ & $7.0\pm^{1.3}_{1.4}$ \\
$\phi^{cb}_b$ & $13.4\pm^{3.0}_{1.4}$ & $18.69\pm^{2.5}_{3.0}$ \\
$\varpi_c-\varpi_b$ & $14.7\pm^{4.0}_{3.3}$ & $20.6\pm^{2.8}_{3.5}$ \\ \hline
$\phi^{be}_b$ & $31.3\pm^{11.9}_{8.7}$ & $47.9\pm^{7.8}_{9.5}$ \\
$\phi^{be}_e$ & $\textup{circulating}$ & $\textup{circulating}$ \\
$\varpi_b-\varpi_e$ & $\textup{circulating}$ & $\textup{circulating}$ \\ \hline
$\phi^{ce}_0$ & $67.5\pm^{25.4}_{19.4}$ & $103.2\pm^{12.3}_{19.2}$ \\
$\phi^{ce}_1$ & $\textup{circulating}$ & $\textup{circulating}$ \\
$\phi^{ce}_2$ & $\textup{circulating}$ & $\textup{circulating}$ \\
$\phi^{ce}_3$ & $\textup{circulating}$ & $\textup{circulating}$ \\
$\varpi_b-\varpi_e$ & $\textup{circulating}$ & $\textup{circulating}$ \\ \hline
$\Laplace$ & $33.0\pm^{12.4}_{9.3}$ & $50.5\pm^{7.9}_{10.0}$ \\